\documentclass[11pt]{article}
\usepackage{epsfig}
\usepackage{graphicx}
\usepackage{citesort}
\usepackage{amsmath,amssymb,amsfonts}
\usepackage{amsmath}
\usepackage{amsfonts}
\usepackage{amssymb}
\usepackage{dsfont,amsthm}
\usepackage[dvips]{feynmp}
\usepackage{slashed}
\usepackage[svgnames,table,hyperref]{xcolor}
\usepackage{graphicx,ifpdf}
\usepackage{upgreek}
\usepackage{floatflt}
\usepackage{amsxtra,mathrsfs,latexsym}\usepackage{dsfont}
\newcommand{\diff}{\text{\rm d}}

\newcommand{\euler}{\text{\rm e}}\newcommand{\imu}{\text{\rm i}}

\newcommand{\field}[1]{\mathds{#1}}

\newcommand{\Z}{\field {Z}}

\usepackage{color}

\ifpdf
\fi
\usepackage[dvips,textwidth=16.0cm,textheight=24.7cm]{geometry}
\numberwithin{equation}{section}

\newcommand{\beq}{\begin{equation}}
\newcommand{\eeq}{\end{equation}}
\newcommand{\bea}{\begin{eqnarray}}
\newcommand{\eea}{\end{eqnarray}}

\definecolor{olive}{rgb}{.5,0.5,0}

\begin{document}
\title{\bf Dynamics and thermodynamics of a \\nonlocal  Polyakov--Nambu--Jona-Lasinio model\\ with running coupling   \footnote{Work supported in part by BMBF, GSI, the DFG Excellence Cluster ``Origin and Structure of the Universe'' and by the Elitenetzwerk Bayern.}}
\author{
T.~Hell, S.~R\"o{\ss}ner, M.~Cristoforetti, and W.~Weise\\
\\{\small Physik-Department, Technische Universit\"at M\"unchen, D-85747 
Garching, Germany}
}
\date {\today}
\maketitle


\begin{abstract}

A nonlocal covariant extension of the two-flavor Nambu and Jona-Lasinio (NJL) model is constructed, with built-in constraints from the running coupling of QCD at high-momentum and instanton physics
at low-momentum scales. Chiral low-energy theorems and basic current algebra relations involving pion properties are shown to be reproduced. The momentum-dependent dynamical quark mass derived from this approach is in agreement with results from Dyson-Schwinger equations and lattice QCD. At finite temperature, inclusion of the Polyakov loop and its gauge invariant coupling to quarks reproduces the dynamical entanglement of the chiral and deconfinement  crossover transitions as in the (local) PNJL model, but now without the requirement of introducing an artificial momentum cutoff. Steps beyond the mean-field approximation are made including mesonic correlations through quark-antiquark ring summations. Various quantities of interest (pressure, energy density, speed of sound etc.) are calculated and discussed in comparison with lattice QCD thermodynamics at zero chemical potential. The extension to finite quark chemical potential and the phase diagram in the $(T,\mu)$-plane are also discussed.

\end{abstract}

\begin{section}{Introduction}

The thermodynamics of strongly interacting quark-gluon matter is a topic of fundamental interest from several perspectives, including the physics of the early universe and ultra-relativistic
heavy-ion collisions. Modeling the phases of QCD in comparison with results from lattice simulations
has become an active part of this field. A promising approach in this direction is the synthesis of the time-honored Nambu and Jona-Lasinio (NJL) model \cite{Nambu,Vogl} with the Polyakov loop \cite{Pisarski}. This so-called PNJL model \cite{Fukushima1,Fukushima2,Ratti1,Simon1,Ratti2,Simon2,Meisinger,Megias} is based on two principal symmetries: the chiral $\text{SU}(N_\text{f})_\text{L}\times \text{SU}(N_\text{f})_\text{R}$ symmetry of QCD with $N_\text{f}$ massless quark flavors, and the symmetry associated with the center $Z(3)$ of the color gauge group. 

The NJL model offers a schematic but nonetheless quite realistic picture of the basic dynamics behind spontaneous chiral symmetry breaking, with the chiral (quark) condensate and pions as Goldstone bosons emerging from a chiral invariant, local four-point interaction between quarks. Confinement is absent in the NJL model. The Polyakov loop, on the other hand, is introduced as the order parameter describing the confinement-deconfinement transition in pure-gauge QCD. In the PNJL model, the minimal gauge invariant coupling of the Polyakov loop to quarks produces a characteristic dynamical entanglement of the chiral and deconfinement transitions. This phenomenon is indeed observed in recent lattice QCD computations \cite{Cheng}, while at the same time its more detailed assessment is still under dispute \cite{Aoki}. Irrespective of such remaining open questions, the qualitative comparison of a variety of PNJL model results with those from lattice QCD has so far turned out to be surprisingly successful.
 
The price paid for the simplicity of the NJL model, with its local point couplings between quarks, is the necessity of introducing a momentum cutoff to regularize loop integrals. The model draws a schematic picture of the QCD interaction by reducing it to a constant at low momenta, $|\vec{p}\,| \lesssim \Lambda_\text{NJL} \sim 0.6$--$0.7$~GeV, and simply turning off the interaction for $|\vec{p}\,| > \Lambda_\text{NJL}$. The applicability of the model is therefore restricted to energy and momentum scales (temperatures, chemical potentials) small compared to $\Lambda_\text{NJL}$. Connections with the running QCD coupling and the established high-momentum, high-temperature behavior governed by perturbative QCD are ruled out right from the start. 

In the present work we develop a nonlocal generalization of the PNJL model with running coupling, capable of describing both chiral and  deconfinement transitions, but without the cutoff of the standard NJL model. This is accomplished by making contact with the Dyson-Schwinger approach to QCD \cite{dse}. Nonlocal extensions of the NJL model are designed to remove deficiencies of the local theory, while at the same time the nonlocal interactions regularize the model in such a way that anomalies are preserved \cite{Ruiz}, charges are properly quantized and currents conserved. The effective interaction is finite to all orders in the loop expansion and, hence, there is no need to introduce extra cutoffs. Our approach takes over those benefits; it differs from other works (see e.\,g.~Refs.~\cite{Ripka,Scoccola,Blaschke,Sasaki,Abuki}) in that we do not use a separable interaction but derive an explicit form for the effective interaction guided by full QCD. The running coupling is constructed in accordance with perturbative QCD at high-momentum scales. The interpolation down to low momenta is performed in close correspondence with the instanton model. Finally, the Polyakov loop is implemented as in the local PNJL model, through the introduction of an additional homogeneous temporal background field with standard $\text{SU}(3)_\text{c}$ gauge invariant coupling to the quarks.

This paper is organized as follows. In Section~\ref{ANJLconstructionSect} we first construct a nonlocal extension of the NJL model at zero temperature. A detailed derivation is given of a nonlocal effective interaction constrained by Dyson-Schwinger and lattice QCD results. The momentum dependent dynamical quark mass is calculated. Pion properties are derived, and it is demonstrated that fundamental chiral symmetry and current algebra relations are properly reproduced. The thermodynamics (restricted  to vanishing chemical potentials) and the implementation of the Polyakov loop is described in Section~\ref{QCDthermoSect}.  The resulting nonlocal PNJL model is applied to calculate the chiral and deconfinement crossover transitions. The outcome is discussed in comparison with lattice QCD results and previous calculations. Pressure and energy density, conformal measure and speed of sound
are computed around the transition temperature, first in mean-field approximation and then with inclusion of mesonic modes (primarily the pion). The extension to finite quark chemical potential is briefly outlined. In Section~\ref{Conclusion} we present our conclusions and an outlook.

\end{section}

\begin{section}{Nonlocal Nambu--Jona-Lasinio Model}\label{ANJLconstructionSect}

\begin{subsection}{Nonlocal NJL action constrained by QCD}\label{nonlocalaction}

Consider as a starting point the basic quark color current, $J^\mu_a = \bar{\psi}(x)\gamma^\mu{\lambda_a\over 2}\psi(x)$, with the quark fields $\psi(x)$ and the generators $\lambda_a/2$ of the $\text{SU}(3)$ color gauge group. The QCD effective interaction between two such color currents contributes a term of the following generic form to the action: 
\begin{equation}\label{Sint}
 \mathcal{S}_\text{int}=-\int\diff^4 x\int\diff^4 y\,J_a^\mu(x)\,\mathcal{G}_{\mu\nu}^{ab}(x-y)\,J_b^\nu(y)\,,
\end{equation}
where $\mathcal{G}_{\mu\nu}^{ab}(x-y)$ is proportional to the gluonic field correlator. Approximate
solutions for this correlator can be obtained from Dyson-Schwinger calculations \cite{dse} and from lattice QCD
\cite{latgluon}. We do not need to specify $\mathcal{G}_{\mu\nu}$ in all detail. We just note that Eq.~\eqref{Sint} defines a nonlocal effective interaction between quarks. The range of the nonlocality is determined by the correlation length characteristic of the color exchange through gluon fields.

Consider next the diagonal parts of $\mathcal{G}_{\mu\nu}$ that have a leading behavior
 $\mathcal{G}_{\mu\nu}^{ab}(x-y) \sim g_{\mu\nu}\,\delta_{ab}\,\mathcal{G}(x-y)$, with a distribution function $\mathcal{G}(x-y)$ of mass dimension 2. Then perform a Fierz exchange transformation  which turns 
$\mathcal{S}_\text{int}$ into a sum of terms
\begin{equation}
 	\mathcal{S}_\text{int} = \sum_\alpha c_\alpha \int\diff^4 x\int\diff^4 y\,\bar\psi(x)\,\varGamma_\alpha\,\psi(y)\,\mathcal{G}(x-y)\,\bar\psi(y)\,\varGamma^\alpha\,\psi(x)\,,
	\label{Fierz1}
\end{equation}
with well-defined Fierz expansion coefficients $c_\alpha$. The $\varGamma_\alpha$ are a set of Dirac, flavor and color matrices, resulting from the Fierz transform, with the property $\gamma_0\,\varGamma_\alpha^\dagger\,\gamma_0=\varGamma_\alpha$.
By a change of variables, Eq.~\eqref{Fierz1} can be rewritten as: 
\begin{equation*}\label{Fierz2}
\mathcal{S}_\text{int}= \sum_\alpha c_\alpha \int\diff^4 x\int\diff^4 z\,\bar\psi\!\left(x+\frac{z}{2}\right)\varGamma_\alpha\,\psi\!\left(x-\frac{z}{2}\right)\,\mathcal{G}(z)\,\bar\psi\!\left(x-\frac{z}{2}\right)\varGamma^\alpha\,\psi\!\left(x+\frac{z}{2}\right).
	\tag{\ref{Fierz1}$'$}
\end{equation*}

Throughout this work we restrict ourselves to the two-flavor case with $\psi(x) = (u(x), d(x))^\top$ and the quark mass matrix $\hat{m} = \text{diag}(m_u, m_d)$. We will work in the isospin-symmetric limit with
$m_u = m_d \equiv m_q$. The Fierz transformation introduces combinations of operators $\varGamma_\alpha$ which maintain the symmetries of the original QCD Lagrangian. The symmetry that governs low-energy QCD with two flavors is global chiral $\text{SU}(2)_\text{L}\times\text{SU}(2)_\text{R}$. A minimal subset of operators satisfying this symmetry is the color-singlet pair of scalar-isoscalar and pseudoscalar-isovector operators, $\varGamma_\alpha =(1,\imu\gamma_5\,\vec{\tau}\,)$, the one we focus on. Other less relevant  operators (vector and axialvector terms in color singlet and color octet channels) will be ignored in the present paper.

Our nonlocal NJL model is thus defined by the following Lagrangian:
\begin{align}
\mathcal{L}(x) &=\bar\psi(x)(\imu\slashed{\partial}-\hat{m})\psi(x) + \mathcal{L}_\text{int}(x)\,,\\ 
\mathcal{L}_\text{int}(x)&=\int\diff^4 z \,\mathcal{G}(z)\left[\bar\psi(x_+)\psi(x_-)\,\bar\psi(x_-)\psi(x_+)+ \bar\psi(x_+)\,\text{i}\gamma_5\,\vec{\tau}\,\psi(x_-)\cdot\bar\psi(x_-)\,\text{i}\gamma_5\,\vec{\tau}\,\psi(x_+)\right]\,,
\nonumber
\end{align}
with $x_\pm = x \pm z/2$. The action in Minkowski space with $x^\mu = (x_0, \vec{x}\,)^\top$ is $\mathcal{S} = \int \diff^4x\,\mathcal{L}$.
From here on and throughout this work calculations will be performed in Euclidean space with the
corresponding action
\begin{equation}
\label{S}
\mathcal{S}_\text{E} = \int \diff^4x\,\mathcal{L}_\text{E}=\int\diff^4 x\,\bar\psi(x)(-\imu\slashed{\partial}+\hat{m})\psi(x)+\mathcal{S}_\text{int}^\text{E}~~,
\end{equation}
where $x = (\vec{x}, x_4)$ etc.~are now Euclidean space-time coordinates.
For convenience we replace the function $\mathcal{G}(z)$ representing the nonlocal interaction kernel by a coupling constant times a normalized distribution,
\begin{equation*}\label{Sa}
\mathcal{G}(z) = {G\over 2}\,\mathcal{C}(z)\,,
\tag{\ref{S}a}
\end{equation*}
with $\int d^4z \,\mathcal{C}(z) = 1$. The coupling strength $G$ of dimension $(\text{length})^2$ (i.e. mass dimension $-2$) is understood to absorb the corresponding Fierz coefficient. Note that the standard (local) NJL model follows for the limiting case
$\mathcal{C}(z)  = \delta^4(z)$.

The bosonization of the Euclidean action is then performed as usual by introducing auxiliary fields $\varphi_\alpha(x)=\left(\sigma(x),\vec{\pi}(x)\right)$, where $\sigma$ and $\vec{\pi}$ are chiral partner boson fields representing scalar-isoscalar and pseudoscalar-isovector mesonic degrees of freedom, respectively.  
Writing the generating functional in  path integral formalism, $Z=\int\mathscr{D}\bar\psi\mathscr{D}\psi\,\exp[-\mathcal{S}_\text{E}]$, we finally obtain 
\begin{equation}
		Z=\int\mathscr{D}\sigma\mathscr{D}\vec{\pi}\,\exp[-\mathcal{S}_\text{E}^\text{bos}]\ ,
\end{equation}
with
\begin{equation}\label{Sbos}
		\mathcal{S}_\text{E}^\text{bos}=-\ln\,\det \hat A+\dfrac{1}{2G}\int\dfrac{\diff^4 p}{(2\pi)^4} \,\phi_\alpha(p)\,\phi_\alpha^*(p)\,,		
\end{equation}
where we have introduced 
\begin{equation*}
\phi_\alpha(p) = \int \diff^4x \,\,\euler^{-\imu p\cdot x}\varphi_\alpha(x)\,.
\end{equation*}
Eq.~\eqref{Sbos} involves the momentum space representation of the operator $\hat A$, as follows: 
\begin{equation}\label{A}
		A(p,p'):=\langle p|\hat A|p'\rangle=\left(-\slashed{p}+m_q\right)(2\pi)^4 \delta^{(4)}(p-p')+\tilde{\mathcal{C}}\!\left(\frac{p+p'}{2}\right)\varGamma_\alpha\, \text{Re}\left[\phi_\alpha(p-p')\right]\,,
\end{equation}
with 
\begin{equation*}
\tilde{\mathcal{C}}(p) = \int \diff^4z \,\,\euler^{-\imu p\cdot x}\,\mathcal{C}(z)\,.
\end{equation*}
We use the notation $\mathcal{C}(p) =: \tilde{\mathcal{C}}(p)$ for simplicity and note that $\mathcal{C}(p=0) = 1$. Details of the  bosonization procedure  are given in Appendix~\ref{bosonizationApp}.

\end{subsection}

\begin{subsection}{Mean field approximation and beyond}

Next, assume that in the homogeneous and isotropic vacuum, the scalar $\sigma$ field has a nonzero expectation value $\bar\sigma = \langle\sigma\rangle$, while the vacuum expectation values of the pseudoscalar fields $\pi_i$ are zero. We write $\sigma(x)=\bar\sigma+\delta\sigma(x),~\vec{\pi}(x)=\delta\vec{\pi}(x)$ and expand the bosonized action \eqref{Sbos} around the mean field in powers of the mesonic fluctuations $\delta\sigma, \delta\vec{\pi}$:
\begin{equation}
 	\mathcal{S}_\text{E}^\text{bos}=\mathcal{S}_\text{E}^\text{MF}+\mathcal{S}_\text{E}^{(2)}+\dots
\end{equation}
The mean field contribution per four-volume $V^{(4)}$ is given by
\begin{equation}\label{SMF}
{\mathcal{S}_\text{E}^\text{MF}\over V^{(4)}}=-4 N_\text{c}\int\dfrac{\diff^4 p}{(2\pi)^4}\ln\left[p^2+M^2(p)\right]+\dfrac{\bar\sigma^2}{2G}\ ,
\end{equation}
with the mass function $M(p)$ determined by the gap equation
\begin{equation*}\label{dynamicalmass}
M(p)=m_q+\mathcal{C}(p)\,\bar\sigma\ .
\tag{\ref{SMF}a}
\end{equation*}
 The quadratic terms beyond mean field approximation are derived explicitly in Appendix~\ref{taylor}. Here we only state the result:
\begin{equation}\label{S2}
\mathcal{S}_\text{E}^{(2)}=\dfrac{1}{2}\int\dfrac{\diff^4 p}{(2\pi)^4}\left[ F^+(p^2)\,\delta\sigma(p)\,\delta\sigma(-p)+F^-(p^2)\,\delta\vec{\pi}(p)\cdot\delta\vec{\pi}(-p)\right],
\end{equation}
where
\begin{equation}\label{G}
	\begin{aligned}
	F^\pm(p^2)&=\dfrac{1}{G}-8N_\text{c}\int\dfrac{\diff^4 q}{(2\pi)^4}\, \mathcal{C}(q)\,\mathcal{C}(q+p)\dfrac{q\cdot (q+p)\mp M(q)M(q+p)}{\big[q^2+M^2(q)\big]\big[\left(q+p\right)^2+M^2(q+p)\big]}\\
	&=\dfrac{1}{G}-
\parbox[height=100pt]{50pt}{\includegraphics{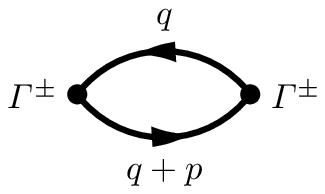}
}\ 
\end{aligned}
\end{equation}
and $\varGamma^+,\varGamma^-$ denote the scalar-isoscalar and pseudoscalar-isovector vertices, respectively. 
 The loop diagram involves the fermion (quark) quasiparticle propagators (in Euclidean space\footnote{We use the convention $p_4:=\imu p_0, \gamma_4:=\imu\gamma_0$.})
\begin{equation}\label{ANJLfermion}
 	S_\text{F}(p)=\quad\parbox[height=200pt]{50pt}{\includegraphics{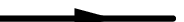}
			}
	\quad=\dfrac{1}{-\slashed{p}+M(p)}
\end{equation}
with the momentum-dependent (dynamical) constituent quark mass $M(p)$ (cf.~Sect.~\ref{politzer}).

\end{subsection}

\begin{subsection}{Gap equation and chiral condensate}
 
The mean field part, $\mathcal{S}_\text{E}^\text{MF}$, of the action $\mathcal{S}_\text{E}$ is  governed by the scalar mean field $\bar\sigma$. Its value is found by the principle of least action, $\frac{\delta \mathcal{S}_\text{E}^\text{MF}}{\delta\sigma}=0$ for $\sigma=\bar\sigma$.
It is straightforward to derive
\begin{equation}\label{gap}
\bar\sigma=8N_\text{c} G\int\dfrac{\diff^4 p}{(2\pi)^4} \,\mathcal{C}(p)\,\dfrac{M(p)}{p^2+M^2(p)}\ ,
\end{equation}
to be solved self-consistently with the gap equation \eqref{dynamicalmass}.
The chiral condensate $\langle\bar\psi\psi\rangle=\langle\bar u u\rangle+\langle \bar d d\rangle$ can be calculated from $\mathcal{S}_\text{E}^\text{bos}$ using the Feynman-Hellmann theorem, by differentiation with respect to the current quark mass $m_q$.

Equivalently, one can use the definition
\begin{equation}
	\langle\bar\psi\psi\rangle=-\imu\,\text{Tr}\,\lim_{y\to x^+}\left[S_\text{F}(x,y)-S_\text{F}^{(0)}(x,y)\right],
\end{equation}
with the full fermion Green function,
\begin{equation}
 	S_\text{F}(x,y)=\int\dfrac{\diff^4 p}{(2\pi)^4}\,\euler^{\imu p(x-y)} S_\text{F}(p)\,,
\end{equation}
and the free quark propagator  $S_\text{F}^{(0)}$ subtracted. Using Eq.~\eqref{ANJLfermion} this leads to
\begin{equation}\label{chiralcondensate}
\langle\bar\psi\psi\rangle=-4 N_\text{f} N_\text{c}\int\dfrac{\diff^4 p}{(2\pi)^4}\left[\dfrac{M(p)}{p^2+M^2(p)}-\dfrac{m_q}{p^2+m_q^2}\right].
	\end{equation}
Note that $M(p)\to m_q$ for large $p$. The subtraction makes sure that no perturbative artifacts are left over in $\langle\bar\psi\psi\rangle$ for $m_q\neq0$.

At this point a comparison with the local NJL model is instructive. There, $\mathcal{C}(p)$ and $M(p)$ are replaced by constants, $\mathcal{C}(0)=1$ and $M$. The relevant momentum space integrations are cut off at $|\vec{p}\,|=\Lambda_\text{NJL}$. The gap equation is simply $M=-G\langle\bar\psi\psi\rangle=\bar\sigma$.  This direct proportionality between the scalar mean field $\bar\sigma$ and the chiral condensate is not realized any more in the nonlocal model, as the inspection of Eqs.~\eqref{gap} and \eqref{chiralcondensate} demonstrates.

\end{subsection}

\begin{subsection}{Pion mass, quark-pion coupling constant and pion decay constant}\label{observables}

 The pion mass $m_\pi$ is determined by the pole of the pion propagator, while the square of the quark-pion coupling constant, $g_{\pi qq}^2$, figures as the residue of the pion pole in the pseudoscalar-isovector quark-antiquark amplitude,
\begin{equation}\label{pionprop}
 	D_{ij}^\pi(q)=\imu\gamma_5\tau_i\dfrac{\imu g_{\pi qq}^2}{q^2+m_\pi^2}\imu\gamma_5\tau_j\,.
\end{equation}
Its inverse is easily calculated from the bosonized action \eqref{Sbos} by means of functional differentiation, i.\,e.
\begin{equation}
	\left[D_{ij}^\pi(q)\right]^{-1}=\dfrac{\delta^2 \mathcal{S}_\text{E}^\text{bos}}{\delta\pi_i(q)\delta\pi_j(0)}=\delta_{ij}F^-(q^2)\,.
\end{equation}
The last equality results from Eqs.~\eqref{S2} and \eqref{G}. The pion mass is therefore determined by
\begin{equation}\label{pionmassdef}
	F^-(-m_\pi^2)=0\,.
\end{equation}
Furthermore, by comparison with Eq.~\eqref{pionprop}  the quark-pion coupling constant is given as:
\begin{equation}\label{gpiqqdef}
	g^{-2}_{\pi qq}=\left.\dfrac{\diff F^-(q^2)}{\diff q^2}\right|_{q^2=-m_\pi^2}.
\end{equation}

The calculation of the pion decay constant $f_\pi$ is more involved. It is defined through the matrix element of the axial current $J^A_{i\mu}(x)$ between the vacuum and the  physical one-pion state $\tilde{\pi}_j$ with $\tilde{\pi}_j=g^{-1}_{\pi qq}\,\pi_j$:
\begin{equation}\label{piondecaydef}
 	\langle0|J^A_{i\mu}(0)|\tilde{\pi}_j(p)\rangle=\imu\delta_{ij} p_\mu f_\pi\,.
\end{equation}
Note that the usual local axial current, $J^{A,\,\text{loc}}_{i\mu}=\bar\psi(x)\gamma_5\frac{\tau_i}{2}\gamma_\mu\psi(x)$, is not conserved within the nonlocal framework (cf.~Ref.~\cite{Bowler}). In order to obtain an explicit expression for the axial current, we gauge the action $\mathcal{S}_\text{E}$ by introducing a set of axial gauge fields $\mathcal{A}^i_\mu(x)$ in the standard way;  partial derivatives are replaced by covariant derivatives, i.\,e.~$\partial_{\mu}\to\partial_{\mu}+\frac{\imu}{2}\gamma_5\,\tau_i \mathcal{A}^i_\mu(x)$, and due to the nonlocal nature of the theory, interaction currents have to be replaced by \cite{Scoccola,Bos,Noguera}
\begin{equation*}
 	\bar\psi\!\left(x+\frac{z}{2}\right)\varGamma_\alpha\psi\!\left(x-\frac{z}{2}\right)\to\bar\psi\!\left(x+\frac{z}{2}\right)\mathcal{W}\!\left(x+\frac{z}{2},x\right) \varGamma_\alpha\,\mathcal{W}\!\left(x,x-\frac{z}{2}\right) \psi\!\left(x+\frac{z}{2}\right),
\end{equation*}
with the function
\begin{equation}\label{comp}
 	\mathcal{W}(x,y)=\mathcal{P}\left\{\exp\left[\frac{\imu}{2}\int_x^y\diff s_\mu\,\gamma_5\, \tau^a \mathcal{A}^a_\mu\right]\right\},
\end{equation}
including the path-ordering operator $\mathcal{P}$. This assures that the gauged action $\mathcal{S}_{\mathcal{A}}$ is invariant with respect to local axial transformations.

After those replacements a gauged bosonized action, $\mathcal{S}_{\mathcal{A}}^\text{bos}$, can be built in the same fashion as outlined in Appendix~\ref{bosonizationApp}. From there it is easy to calculate the matrix element required for the determination of $f_\pi$ according to Eq.~\eqref{piondecaydef}:
\begin{equation*}
 	\langle0|J^A_{i\mu}(0)|\tilde\pi_j(p)\rangle=\left.\dfrac{\delta^2 \mathcal{S}_{\mathcal{A}}^\text{bos}}{\delta\tilde\pi^j(p)\,\delta\mathcal{A}^{i\mu}}\right|_{\mathcal{A}=0}.
\end{equation*}
Finally, switching to momentum space and evaluating the functional derivatives and traces according to Appendix~\ref{taylor}
 leads to the pion decay constant in mean field approximation\footnote{Note, that the result stated here originates from an RPA calculation with vertex functions chosen such that Ward identities are preserved (see Ref.~\cite{Bowler}). The complete result, however, is more involved and its derivation is relegated to Appendix~\ref{decayconstant}. The  result \eqref{piondecay} corresponds to the leading order expansion in $m_\pi$ of \eqref{pionapp}.} (see~also Ref.~\cite{Scoccola}):
\begin{equation}\label{piondecay}
	f_\pi=g_{\pi qq}\dfrac{\mathcal{F}(-m_\pi^2)-\mathcal{F}(0)}{m_\pi^2}\ ,
\end{equation}
where we have defined
\begin{equation*}
 	\mathcal{F}(p^2)=m_q\, \mathcal{I}_1(p^2)+\bar\sigma\, \mathcal{I}_2(p^2)\,,
\end{equation*}
with
\begin{equation*}
	\begin{aligned}
 	\mathcal{I}_1&=8 N_\text{c}\int\dfrac{\diff^4q}{(2\pi)^4}\,\mathcal{C}(q)\dfrac{q\cdot (q+p)+M(q)M(q+p)}{\left[q^2+M^2(q)\right]\left[(q+p)^2+M^2(q+p)\right]}\\
	\mathcal{I}_2&=8 N_\text{c}\int\dfrac{\diff^4q}{(2\pi)^4}\,\mathcal{C}(q)\,\mathcal{C}(q+p)\dfrac{q\cdot (q+p)+M(q)M(q+p)}{\left[q^2+M^2(q)\right]\left[(q+p)^2+M^2(q+p)\right]}
\ .
	\end{aligned}
\end{equation*}
Two pion observables in vacuum, $f_\pi$ and $m_\pi$, together with $g_{\pi qq}$ will be used to fix the parameters of the nonlocal model of this work (see Sect.~\ref{parameterfitSect}).

\end{subsection}

\begin{subsection}{Chiral relations}

In order to confirm that our model is consistent with fundamental chiral symmetry requirements, we derive the Goldberger--Treiman and Gell-Mann--Oakes--Renner relations. For this purpose it is sufficient to consider an expansion of the integrals $\mathcal{I}_1,\mathcal{I}_2$ up to $\mathcal{O}(m_\pi^2)$, i.\,e. to leading (linear) order in the quark mass $m_q$. First note that using Eqs.~\eqref{gap} and \eqref{pionmassdef}, one can write $\mathcal{F}(0)=\bar\sigma/G$ and $\mathcal{I}_2(-m_\pi^2)=1/G$. Then one obtains from Eq.~\eqref{piondecay}
\begin{equation}\label{mpifpi}
 	m_\pi^2 f_\pi=m_q\,g_{\pi qq}\,\mathcal{I}_1(-m_\pi^2)\,.
\end{equation}
A Taylor expansion about the chiral limit of the pion polarization term, Eq.~\eqref{G}, leads to
\begin{equation*}
 	F^-(p^2)=F^-(0)+\left.\dfrac{\partial F^-}{\partial p^2}\right|_{p^2=0} p^2+\mathcal{O}(p^4)=F^-(0)+g_{\pi qq}^{-2}\, p^2+\mathcal{O}(p^4)\,.
\end{equation*}
Using this in the conditional equation for the pion mass, $F^-(-m_\pi^2)=0$, one recovers the leading order contribution to $m_\pi^2$:
\begin{equation}\label{taylorpi}
 	m_\pi^2=\dfrac{g_{\pi qq}^2}{\bar\sigma} m_q \mathcal{I}_1+\mathcal{O}(m_\pi^4)=g_{\pi qq}^2\, F^-(0)+\mathcal{O}(m_\pi^4)\,.
\end{equation}
Finally, combining Eqs.~\eqref{G} and \eqref{chiralcondensate} gives
\begin{equation*}
 	F^-(0)=-\dfrac{m_q}{\bar\sigma^2}\langle\bar\psi\psi\rangle+\mathcal{O}(m_q^2)\,.
\end{equation*}
From Eq.~\eqref{mpifpi} and the first equality of Eq.~\eqref{taylorpi} one finds
\begin{equation}\label{gt}
 	f_\pi\, g_{\pi qq}=\bar\sigma+\mathcal{O}(m_q)\,,
\end{equation}
which, in the chiral limit, i.\,e.~for $m_q=0$, is nothing but the Goldberger--Treiman relation at the level of quarks as quasiparticles. Using this relation, the second equality of Eq.~\eqref{taylorpi} together with the expression for $F^-(0)$ derived above  gives
\begin{equation}\label{gmor}
 	m_\pi^2 f_\pi^2=-m_q\langle\bar\psi\psi\rangle+\mathcal{O}(m_q^2)\,,
\end{equation}
which is the well-known Gell-Mann--Oakes--Renner relation. We have thus demonstrated that the nonlocal NJL model preserves chiral low-energy theorems and current algebra relations. 

\end{subsection}

\begin{subsection}{Determination and fixing of the distribution {\boldmath{$\mathcal{C}(p)$}}}\label{politzer}

The Fourier transform $\mathcal{C}(p)$ of the nonlocality distribution introduced in section \ref{nonlocalaction} is a key quantity and the basic input of the present approach. QCD constraints are used to determine $\mathcal{C}(p)$, both in the high-$p$ (perturbative) and low-$p$ (non-perturbative) regions, as follows.

\begin{subsubsection}{Quark self-energy at large momentum}
 
Consider a quark propagating with large (Euclidean) momentum $p$ in the QCD vacuum. Its self-energy $\varSigma(p)$ is given pictorially as
\begin{equation}\label{fulldiagram}
 -\imu\varSigma(p)=\  
\parbox{120pt}{\includegraphics{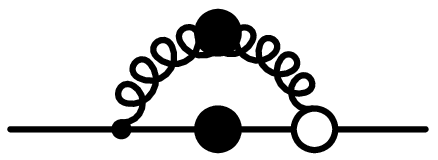}
}\ ,
\end{equation}

\vspace{-.75cm}

\noindent with full quark and gluon propagators and vertex functions\footnote{In a self-consistent Schwinger-Dyson approach the filled circles denote fully dressed propagators and the open circle is the irreducible quark-gluon vertex.}. At high momentum, this self-energy can be evaluated recalling its operator product expansion and identifying the leading $\mathcal{O}(p^{-2})$ term \cite{Politzer,Miransky}.

It is well known that QCD perturbation theory at any order does not generate a non-zero quark mass term starting from a massless quark, simply as a consequence of helicity conservation at each quark-gluon vertex. However, as pointed out by Politzer \cite{Politzer}, the presence of a non-perturbative QCD vacuum with a quark condensate $\langle\bar\psi\psi\rangle\neq0$ turns a massless current quark into a quasiparticle with a momentum dependent mass (often referred to as a constituent quark): the spontaneous breaking of chiral symmetry implies a non-zero dynamical quark mass.

At high quark momentum, the leading non-trivial part of $\varSigma(p)$ is found replacing the full quark-gluon vertex function by the bare vertex $\gamma_\nu$ (``rainbow truncation'') and replacing the full gluon propagator by its free part, i.\,e.~$4\pi\alpha_\text{s}(p^2) D_{\mu\nu}(p)\to 4\pi\alpha_\text{s}(p^2)D_{\mu\nu}^\text{free}(p)$ (``Abelian approximation''), with the running QCD coupling $\alpha_\text{s}(p^2)$. The full quark propagator in \eqref{fulldiagram} is kept with its non-perturbative pieces. One then arrives at the quark self-energy to leading order in $1/p^2$ (with $p^2>0$):
\begin{equation}\label{quarkselfen}
\varSigma(p^2)=\pi\dfrac{N_\text{c}^2-1}{2 N_\text{c}^2N_\text{f}}\dfrac{\alpha_\text{s}(p^2)}{p^2}(3+\xi)\langle\bar\psi\psi\rangle+\delta\varSigma\,,
\end{equation}
where we have explicitly written the non-perturbative piece, $\varSigma^\text{n.\,p.}$, and the remaining $\delta\varSigma$ stands for all perturbative corrections to $\varSigma$ (which vanish in the chiral limit, $m_q\to0$). The parameter $\xi$ is required for gauge fixing (see also Refs.~\cite{Miransky,McMullan}).

From Eq.~\eqref{quarkselfen} one arrives at the dynamically generated quark mass term
\begin{equation}\label{politzermass}
M(p^2):=\varSigma^\text{n.\,p.}(p^2)=-\pi\dfrac{N_\text{c}^2-1}{2 N_\text{c}^2 N_\text{f}}\dfrac{\alpha_\text{s}(p^2)}{p^2} (3+\xi)\langle\bar\psi\psi\rangle\ .
\end{equation}
This result coincides with the constituent quark mass calculated by Politzer. As mentioned, even in the limit of vanishing current quark masses, the quark propagator has a non-perturbative contribution which generates a large constituent quark mass. In what follows we choose the Landau gauge, setting $\xi=0$. Any other value of $\xi$ can be absorbed by a rescaled coupling in Eq.\,\eqref{politzermass}.

From Eqs.~\eqref{Sa} and \eqref{ANJLfermion} it is evident that the distribution $\mathcal{C}(p)$ must behave as $M(p)$ at large $p$, up to normalization. In order to reproduce the QCD result \eqref{politzermass} in the chiral limit (with $N_\text{c}=3$ and $N_\text{f}=2$), we therefore choose
\begin{equation}
 	\mathcal{C}(p^2)\propto \dfrac{2\pi}{3}\dfrac{\alpha_\text{s}(p^2)}{p^2}\qquad(p\ge\varGamma)\ ,
\end{equation}
at $p$ larger than a matching scale $\varGamma$ of order $1\,\text{GeV}$ below which $\mathcal{C}(p^2)$ is governed by non-perturbative physics (see the following subsection). For our present purposes it is sufficient to use the standard form\footnote{The more precise NNLO form for $\alpha_\text{s}(p^2)$ can of course be used but the resulting changes are insignificant within the accuracy of the present approach.}
\begin{equation}
 	\alpha_\text{s}(p^2)=\dfrac{4\pi}{\beta_0\ln\frac{p^2}{\Lambda^2_\text{QCD}}}
\end{equation}
with $\beta_0=9$ (using three active quark flavors) and $\Lambda_\text{QCD}=0.25\,\text{GeV}$ in order to reproduce $\alpha_\text{s}=0.12$ at $m_Z=91.2\,\text{GeV}$.

\end{subsubsection}

\begin{subsubsection}{Aspects of non-perturbative QCD}
 
The behavior of $\mathcal{C}(p)$ at low momentum $p$, below the matching scale $\varGamma$, is not controlled by first principles and needs guidance e.\,g.~from lattice QCD and other non-perturbative considerations.
A useful scheme for extrapolating to low momenta is provided by the Instanton Model (IM) \cite{Schaefer}. Instantons are non-perturbative gauge configurations representing extrema of the Yang-Mills action. They are known to play an important role in the Euclidean partition function. In particular, it has been demonstrated that instantons provide a mechanism for chiral symmetry breaking in the QCD vacuum. Detailed computations in the context of the Interacting Instanton Liquid Model \cite{Cristoforetti} have shown that, in the chiral regime of QCD, there is not only qualitative but also quantitative agreement between the IM and full QCD, at least for $N_\text{f}=2$. It is also known \cite{Schaefer} that the partition function of the IM can be represented in terms of a nonlocal fermionic effective action similar to that of the nonlocal NJL model,  where now the nonlocality originates in the finite size of the instantons ($\sim 1/3\,\text{fm}$) which mediate the interaction.

The distribution $\mathcal{C}_I(p)$ extracted from the IM is expressed in terms of Bessel functions as \cite{Schaefer}
\begin{equation}\label{regim}
	\mathcal{C}_I(p)=\pi p^2 d^2 \dfrac{\diff}{\diff x}(I_0(x)K_0(x)-I_1(x)K_1(x))\,,\qquad x=\frac{|p| d}{2}\ ,
\end{equation}
where $d\simeq0.35\,\text{fm}$ is the typical instanton size. The implementation of this instanton inspired $\mathcal{C}_I(p)$ in finite-temperature calculations is numerically expensive. For practical convenience we approximate $\mathcal{C}(p)$ at low momentum $p$ by a Gaussian,
\begin{equation}\label{reggauss}
 	\mathcal{C}(p^2)=\exp\left[-\dfrac{p^2 d^2}{2}\right]\qquad{p<\varGamma}\,,
\end{equation}
subject to the requirement that this $\mathcal{C}(p)$ smoothly continues into the logarithmic high-momentum behavior at the matching scale $\varGamma$. The width of the Gaussian ($d^{-1}\simeq0.56\,\text{GeV}$) is then directly determined by the instanton size.

Figure~\ref{CvsCI} shows a comparison of the Gaussian form \eqref{reggauss} and the IM result \eqref{regim} for $\mathcal{C}(p)$, demonstrating that the two curves are indeed quite similar. One should note in addition that, when implementing $\mathcal{C}(p)$ in the gap equation \eqref{gap} determining the dynamical quark mass $M(p)$, or in the calculation of the chiral condensate \eqref{chiralcondensate}, the results are insensitive to the detailed behavior of $\mathcal{C}(p)$ or $M(p)$ at very low momentum. The reason is that the Euclidean momentum space integrals \eqref{gap} and \eqref{chiralcondensate} introduce a measure proportional to $\diff p\,p^3$ so that the maximum weight under those integrals is concentrated in a characteristic window around $p\sim0.8\,\text{GeV}$.

\end{subsubsection}

\end{subsection}

\begin{subsection}{Parameter fixing and dynamical quark mass {\boldmath{$M(p)$}}}\label{parameterfitSect}

Given the nonlocality distribution
\begin{equation}
 	\mathcal{C}(p^2)=\begin{cases} \euler^{-p^2 d^2/2}&\text{for $p^2<\varGamma^2$}\\
 	                  		\text{const.}\cdot\dfrac{\alpha_\text{s}(p^2)}{p^2}&\text{for $p^2\ge\varGamma^2$}\,,
 	                 \end{cases}
\end{equation}
normalized as $\mathcal{C}(p=0)=1$ and with a constant fixed by the matching condition at $p=\varGamma$, there remain only two parameters apart from the matching scale $\varGamma$: the coupling strength $G$ in Eq.~\eqref{Sa} and the (current) quark mass $m_q$ as a measure of explicit chiral symmetry breaking.

The prime advantage of the present nonlocal approach is the absence of an artificial momentum space cutoff as it appears in local NJL type models. Indeed, the gap equation \eqref{gap} does not require any regularization: the scalar mean field $\bar\sigma$ and, likewise, the dynamical quark mass $M(p)$ are well-defined within the nonlocal scheme.\footnote{This does not exclude the possibility that secondary quantities, such as the chiral condensate, can be weakly divergent and potentially require a cutoff at ultrahigh momenta, very far beyond the range of applicability of the model. Inspection of the integral for $\langle\bar\psi\psi\rangle$ in Eq.~\eqref{chiralcondensate}, with the asymptotic form $M(p)\to m_q+\frac{\text{const.}}{p^2\ln p^2}$ inserted, displays a weak (double-logarithmic) far-ultraviolet divergence. For regularization we choose a cutoff at $20\,\text{GeV}$, the necessity of which just reflects the simple leading-order choice for the asymptotics of $\mathcal{C}(p)$.}

The parameters and resulting values of observables are listed in Table~\ref{parameterfix}. The determination of $G$ and $m_q$ is such as to reproduce, as closely as possible, the empirical pion decay constant $f_\pi=92.4\,\text{MeV}$ and the pion mass $m_\pi$. Fig.~\ref{sigmarunning}  shows the momentum dependence of the dynamical quark mass $M(p)$ compared with lattice data from Ref.~\cite{Bowman}. One notes that the nonlocal NJL model reproduces very precisely both the low- and the high-momentum behavior. 

The optimal matching scale $\varGamma\simeq0.8\,\text{GeV}$ turns out to be located in the momentum window where the gap equation \eqref{gap} has its maximum weight. It is therefore important to examine the sensitivity of the results with respect to variations of $\varGamma$. We have performed this test by varying the matching scale in the range $0.6\,\text{GeV}<\varGamma<1\,\text{GeV}$ (i.\,e. about 20\,\% around its optimal value), with the constraint of keeping the chiral condensate $\langle\bar\psi\psi\rangle$ fixed. The resulting variations in the pion mass and decay constant are within only 10\,\%, implying stable conditions.

Note that a direct comparison to the NJL model is not appropriate, because the NJL model uses a three-momentum cutoff for the regularization of the integrals while in the nonlocal approach the distribution $\mathcal{C}(p)$ mimics a four-momentum cutoff. This implies, in particular, a different scale for the coupling strength $G$ appearing in the scalar mean field $\bar\sigma$. Consider again the chiral limit and compare Eqs.~\eqref{gap} and \eqref{chiralcondensate}. One can write the dynamical quark mass as $M(p)=\mathcal{C}(p)\bar\sigma=-\mathcal{C}(p)\bar{C} G\langle\bar\psi\psi\rangle$ where $\bar C=-\int\diff^4 p\,\mathcal{C}(p) f(p)\big/\int\diff^4 p\, f(p)$ with $f(p)=M(p)/\left(p^2+M^2(p)\right)$. Given that the maximum weight under the integrals is located around $p\sim 0.8\,\text{GeV}$, one finds $\bar C\sim\frac{1}{4}$ and, consequently, $M(0)\simeq-\frac{G}{4}\langle\bar\psi\psi\rangle$. Hence with the nonlocality profile as shown in Fig.~\ref{CvsCI}, the coupling $G$ in the present approach should be about four times that of the standard NJL model, which is indeed the case.

\begin{figure}[t]
\begin{center}
	\begin{minipage}[t]{.475\textwidth}{
		\includegraphics[width=\textwidth]{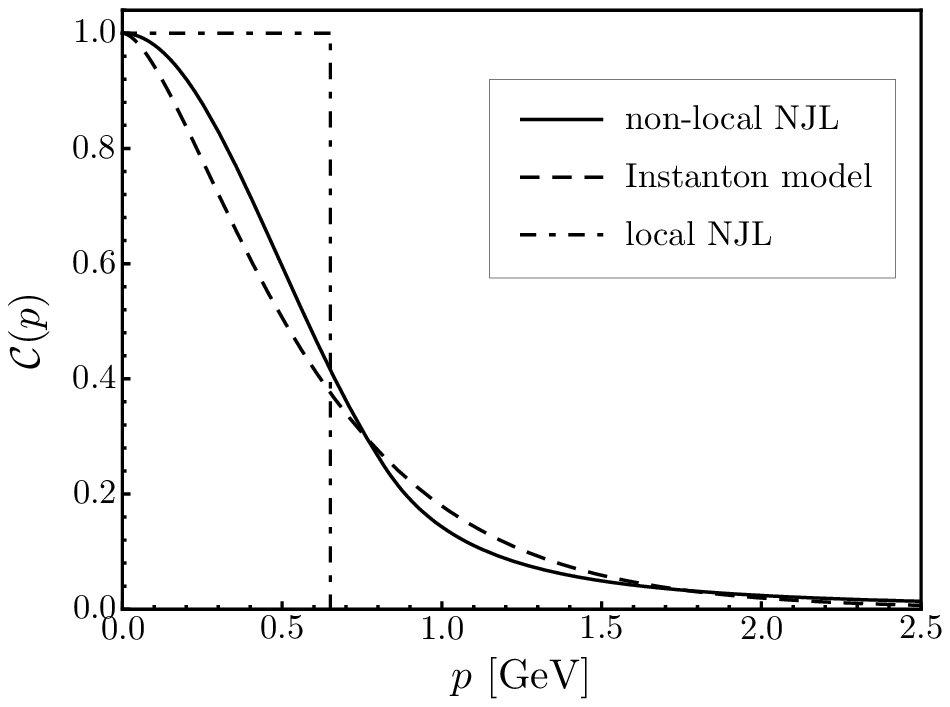}
		\caption{Comparison of the form factor $\mathcal{C}$ used in the nonlocal NJL model (solid line) and the analogous form factor obtained from the Instanton Model (dashed line). The dot-dashed line shows the function $\mathcal{C}(p)$ in an NJL model with four-momentum cutoff.\label{CvsCI}}}
	\end{minipage}
\hfill
	\begin{minipage}[t]{.475\textwidth}{
		\includegraphics[width=\textwidth]{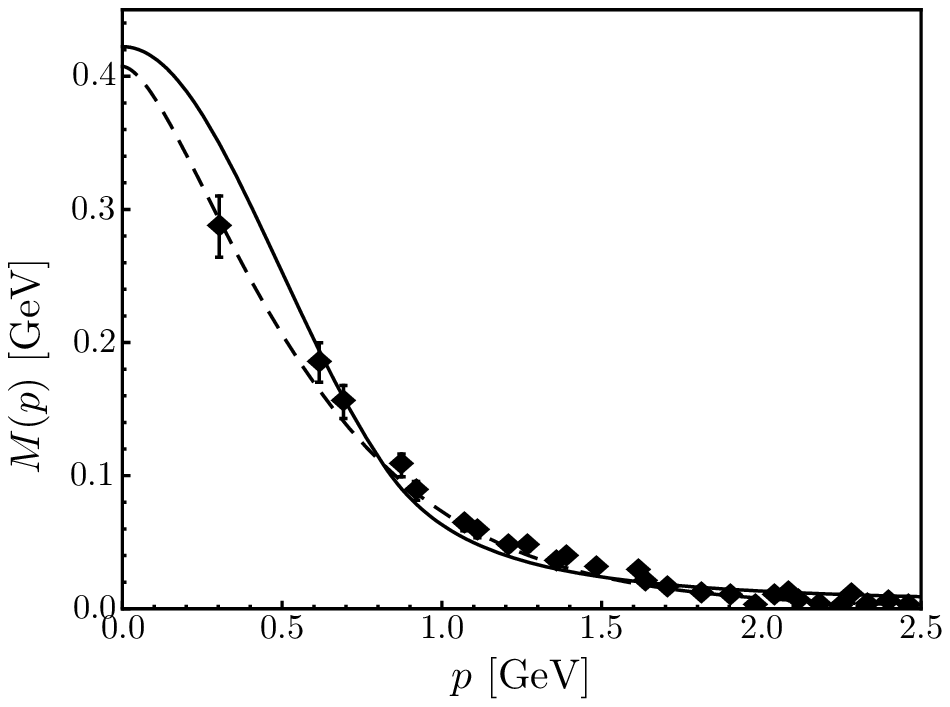}
		\caption{Momentum dependence of the constituent quark mass  $M$ compared with lattice data extrapolated to the chiral limit (from Ref.~\cite{Bowman}). The dashed line shows the mass $M$ calculated fully within the IM for the chiral limit \cite{Schaefer}.\label{sigmarunning}}}
	\end{minipage}
\end{center}
\end{figure}

The quark mass $m_q\simeq3.3\,\text{MeV}$ is compatible with QCD estimates at a typical renormalization scale $\mu\simeq 2\,\text{GeV}$ \cite{particleproperties}. The value of the chiral condensate follows consistently from the Gell-Mann, Oakes, Renner relation \eqref{gmor}, with $\langle\bar u u\rangle=\langle\bar dd\rangle\simeq-(0.28\,\text{GeV})^3$. The matching scale $\varGamma\simeq0.85\,\text{GeV}$ for the low- and high-momentum parts of $\mathcal{C}(p)$ can be considered ``natural''.

\begin{center}
\begin{table}
\begin{tabular}{|c|c|c|}\hline\hline
$m_q$ [MeV]& $\varGamma$ [GeV]& $G$ [GeV$^{-2}$]\\\hline
3.3& 0.83& 41.1\\\hline\hline
\end{tabular}
\qquad
\begin{tabular}{|c|c|c|c|}\hline\hline
$\langle\bar\psi\psi\rangle^{1/3}$ [GeV]& $\bar\sigma$ [GeV]& $f_\pi$ [GeV] & $m_\pi$ [GeV]\\\hline
$-0.36$ & 0.42&0.09&0.14\\\hline\hline
\end{tabular}
\caption{Parameters and calculated physical quantities for the nonlocal NJL model with $N_\text{c}=3$ and $N_\text{f}=2$.}
\label{parameterfix}\end{table}
\end{center}
 
\end{subsection}

\end{section}

\vspace{-1cm}
\begin{section}{Thermodynamics of the nonlocal NJL Model}\label{QCDthermoSect}

In this section we apply the nonlocal NJL approach in order to model QCD thermodynamics. Of primary interest are the chiral and  confinement-deconfinement transitions. This requires coupling the quark quasiparticles to the color gauge fields. One-gluon exchange degrees of freedom (at the level of the quark self-energy) are effectively incorporated in the high-momentum part of $\mathcal{C}(p)$, while chiral dynamics is implemented through the scalar field $\bar\sigma$ which plays the role of a chiral order parameter. The thermodynamics of deconfinement, on the other hand, is introduced by coupling the quarks to the Polyakov loop as the order parameter of the confinement-deconfinement phase transition in pure-gauge QCD. This is the step to the Polyakov-loop-extended NJL (or PNJL) model \cite{Fukushima1,Fukushima2,Ratti1,Simon1,Ratti2,Simon2,Meisinger,Megias}. We now proceed to develop the nonlocal generalization of this PNJL model.

\begin{subsection}{Polyakov loop}

The so-called renormalized Polyakov loop, $\langle\Phi\rangle$, is defined through
\begin{equation}
	\langle\Phi(\vec{x}\,)\rangle=\dfrac{1}{N_\text{c}}\langle\text{tr}_\text{c}\left[L(\vec{x}\,)\right]\rangle\,,
\end{equation}
where $\text{tr}_\text{c}$ denotes the trace over color space only and $L$ is the  Polyakov loop, a timelike Wilson line connecting two points at $t=0$ and $t=-\imu\beta$ in imaginary time space, with periodic boundary conditions:
\begin{equation}
	L(\vec{x}\,)=\mathcal{P}\exp\left\{\imu\int_0^\beta\diff\tau\, A_4(\tau,\vec{x}\,)\right\}\ .
\end{equation}
Here $\mathcal{P}$ is the path-ordering operator and $A_4:=A_4^a t_a=\imu A_0^a t_a$ is the fourth component (in Euclidean space) of the gluon fields\footnote{The $\text{SU}(3)$ matrices $t_a,a\in\{1,\dots,8\}$ are defined through the Gell-Mann matrices $\lambda_a,a\in\{1,\dots,8\}$ as $t_a:=\frac{\lambda_a}{2}$.}.

As shown e.\,g. in Ref.~\cite{McLerran},  $\langle\Phi\rangle=0$ implies an infinite free energy,  corresponding to confinement, while $\langle\Phi\rangle=1$ implies a vanishing free energy of the  system and, thus, deconfinement. According to  Ref.~\cite{Weiss}, $\langle\Phi\rangle$ serves strictly as an  order parameter for the confinement-deconfinement phase transition only in the case where no (not even massless) quarks are present. In  other cases $\langle\Phi\rangle$ is still useful as an indicator for a rapid crossover transition.

\end{subsection}

\begin{subsection}{Coupling of quarks and Polyakov loop}

Gluons can be introduced into the nonlocal NJL model by the standard minimal gauge coupling procedure, substituting the normal  derivate $\partial_\mu$ by the covariant derivative\footnote{The coupling strength $g$ is assumed to be included already in the definition of the gluon fields, i.\,e.~$g A_\mu\to A_\mu$.} $D_\mu=\partial_\mu-\imu A_\mu$ or, in momentum space, $p_\mu\to p_\mu+A_\mu$.\footnote{This minimal substitution procedure can be derived rigorously as follows: the (gluon) gauge fields are introduced in the nonlocal model according to Eq.~\eqref{comp}, with $\gamma_5\tau^a\mathcal{A}_\mu^a$ replaced by $t^a A_\mu^a$ in the exponential. Limiting ourselves to the case of constant gauge fields $A^3_4, A^8_4$, the path-integrations in the exponential can be carried out, generating terms $\euler^{\imu x_4 A^i_4}$. In momentum space, these terms translate into delta functions $\delta(q_4-(p_4-A^i_4))\,\delta^{(3)}(\vec{q}-\vec{p}\,)$ leading to the substitution prescription appropriate for the nonlocal approach.} 

As usual we introduce explicitly particle and antiparticle degrees of freedom switching to Nambu-Gor'kov space. This is particularly useful for the correct gauging of the theory. The fermion propagator is derived from the Euclidean action $\mathcal{S}_\text{E}$, Eq.~\eqref{S}, using standard  functional calculus. The inverse  propagator is given by the fermion determinant already calculated in Eq.~\eqref{A}. Hence, in mean field approximation we have
\begin{equation}\label{Sfermi}
	S^{-1}_\text{MF}(p)=-\slashed{p}+m_q+\mathcal{C}(p)\bar\sigma\,.
\end{equation}
The transition to Nambu-Gor'kov space consists in the introduction of the field $\Psi:=(\psi,\psi^C)^\top$, where $\psi^C=C\bar\psi^\top$ with the charge conjugation matrix $C=\imu\gamma_0\gamma_2$, and the identification
\begin{equation}
	\mathscr{L}=\dfrac{1}{2}\bar\Psi\begin{pmatrix} S^{-1}&0\\ 0& \left[ S^{-1}\right]^C\end{pmatrix}\Psi+\dfrac{G}{8}\left(\bar\Psi\hat\varGamma\Psi\right)^2,
\end{equation}
with $\hat\varGamma=\left(\begin{smallmatrix}\varGamma&0\\0&\varGamma^C\end{smallmatrix}\right)$, $\varGamma^C=-C\varGamma C$.

The introduction of the Nambu-Gor'kov propagator is particularly convenient when finite temperatures are considered. After  bosonization of the Lagrangian the partition function of the theory becomes (with $\beta=1/T$ the inverse temperature):
\begin{equation}\label{Zbos}
	Z=\int\mathscr{D}\varphi_\alpha\int\mathscr{D} A\,\euler^{\ln\,\widetilde{\det}\tilde{A}}\exp\left\{-\int_0^\beta\diff \tau\int\diff^3 x\,\dfrac{\varphi_\alpha(x)\varphi_\alpha^*(x)}{2G}\right\},
\end{equation}
where $\varphi_\alpha$ represents  the mesonic degrees of freedom and $\tilde A=S_\text{F,gauge}^{-1}$ is the Fermi propagator coupled to the gauge fields,  obtained from Eq.~\eqref{A} after the substitution $p_\mu\to p_\mu+A_\mu$.

Without loss of generality the Polyakov loop field can be represented in terms of the $a=3$ and $a=8$ color components of the gluon fields, i.\,e.~we consider only the diagonal matrices of the $\text{SU}_\text{c}(3)$ Lie algebra (the so-called Cartan sub-algebra of $\text{SU}_\text{c}(3)$)  and neglect spatial fluctuations at this stage so that only the time component is relevant:
\begin{equation*}
	A_\mu=\delta_{\mu4}(A_4^3 t_3+A^8_4 t_8)\,.
\end{equation*}
The group integration over the remaining non-diagonal elements gives  the Haar volume 
\begin{equation}\label{JHaar}
	J(\phi_3,\phi_8)=\dfrac{1}{V_{\text{SU}(3)}}\int\prod_{i\in\{1,2,4,5,6,7\}}\diff \phi_i=\dfrac{2}{3\pi^2}\left(\cos(\phi_3)-\cos(\sqrt{3}\phi_8)\right)^2\sin^2(\phi_3)\,,
\end{equation}
where we have set $\phi_{3,8}=\beta \frac{A_4^{3,8}}{2}$.
This volume can be written in terms  of the Polyakov loop\footnote{From now on we omit angled brackets for notational simplicity, i.\,e.~we write $\langle\Phi\rangle\to\Phi$.} $\Phi$ and its complex conjugate $\Phi^*$,
\begin{equation}
 	J(\Phi,\Phi^*)=\dfrac{9}{8\pi^2}\left[1-6\Phi^*\Phi+4\left({\Phi^*}^3+\Phi^3\right)-3\left(\Phi^*\Phi\right)^2\right],
\end{equation}
with $\Phi=\frac{1}{N_\text{c}}\text{tr}_\text{c}\left[\exp\left(\imu(\phi_3\lambda_3+\phi_8\lambda_8)\right)\right]$.  With the partition function \eqref{Zbos} this procedure leads to  the construction of the following effective potential $\mathcal{U}$ that includes the effects of the six non-diagonal gluon fields:
\begin{equation}\label{polyakovU}
	\dfrac{\mathcal{U}(\Phi,\Phi^*,T)}{T^4}=-\dfrac{1}{2} b_2(T)\Phi^*\Phi+b_4(T)\ln\left[1-6\Phi^*\Phi+4\left({\Phi^*}^3+\Phi^3\right)-3(\Phi^*\Phi)^2\right]\,.
\end{equation}
The coefficients are parametrized as
\begin{equation*}\begin{aligned}b_2(T)&=a_0+a_1\left(\dfrac{T_0}{T}\right)+a_2\left(\dfrac{T_0}{T}\right)^2+a_3\left(\dfrac{T_0}{T}\right)^3\\
b_4(T)&=b_4\left(\dfrac{T_0}{T}\right)^3.
\end{aligned}
\end{equation*}
The first term on the right-hand side is reminiscent of a Ginzburg-Landau approach. The values of the coefficients are taken from Ref.~\cite{Simon1} and listed  in Table~\ref{polyakovparameters}. The parametrization \eqref{polyakovU} of the Polyakov loop effective potential $\mathcal{U}$ is applicable at temperatures $T$ up to about twice the critical $T_c$. At higher temperatures, transverse gluon degrees of freedom -- not covered by the Polyakov loop -- begin to be important. Other parametrizations of $\mathcal{U}$ are possible. An example is the two-parameter ansatz \cite{Fukushima3} based on the strong coupling limit. These two versions of $\mathcal{U}$ differ at high temperatures but produce very similar pressure profiles \cite{Fukushima3} at $T\lesssim 2 T_c$, the temperature region of primary interest in the present study.
\begin{table}
\begin{center}
\begin{tabular}{|c|c|c|c|c|c|}
\hline\hline
$a_0$&$a_1$&$a_2$&$a_3$&$b_4$\\\hline\hline
$3.51$&$-2.56$&$15.2$&$-0.62$&$-1.68$\\\hline\hline
\end{tabular}
\caption{Parameters of the Polyakov potential $\mathcal{U}$ (from Ref.~\cite{Simon1}).}\label{polyakovparameters}
\end{center}
\end{table}

We now continue with the evaluation of the partition function \eqref{Zbos}. In the following we first limit ourselves to the mean field approximation. The integrals in Eq.~\eqref{Zbos} are easily performed. The thermodynamic potential $\Omega=-\frac{T}{V}\ln Z$ becomes
\begin{equation}\label{Omega}
	\Omega=-\ln\,\widetilde{\det}\left[\beta\tilde S^{-1}\right]+\dfrac{\bar\sigma^2}{2G}+\mathcal{U}(\Phi,\Phi^*,T)\,,
\end{equation}
where $\tilde S^{-1}$ is the inverse quark propagator expressed in Nambu-Gor'kov space and given in Eq.~\eqref{nambugorkovS}. The functional determinant is evaluated using the Matsubara imaginary time formalism (see e.\,g.~Ref.~\cite{Kapusta}), replacing $p_0\to\imu\omega_n$, with the fermionic Matsubara frequencies $\omega_n=(2n+1)\pi T,(n\in\Z)$, and performing integrals over four-momentum space according to
\begin{equation}\label{matsubarasubs}
	\text{(zero temperature)}\qquad\int\dfrac{\diff^4 p}{(2\pi)^4}\quad\to\quad T\sum_{n\in\Z}\int\dfrac{\diff^3 p}{(2\pi)^3}\qquad\text{(finite temperature)\,.}
\end{equation}
The resulting thermodynamic potential is\footnote{Note an extra factor $\frac{1}{2}$ because of the doubling of the degrees of freedom in Nambu-Gor'kov space.}:
\begin{equation}\label{OmegaMatsubara}
	\Omega=-\dfrac{T}{2}\sum_{n\in\Z}\int\dfrac{\diff^3 p}{(2\pi)^3}\,\text{tr}\,\ln\left[\beta\tilde S^{-1}(\imu\omega_n,\vec{p}\,)\right]+\dfrac{\bar\sigma^2}{2G}+\mathcal{U}(\Phi,\Phi^*,T)\,,
\end{equation}
with
\begin{equation}\label{nambugorkovS}
	\tilde S^{-1}(\imu\omega_n,\vec{p}\,)=\begin{pmatrix}\imu\omega_n\gamma_0-\vec{\gamma}\cdot\vec{p}-\tilde M-\imu A_4\gamma_0&0\\ 0&\imu\omega_n\gamma_0-\vec{\gamma}\cdot\vec{p}-\tilde M^*+\imu A_4\gamma_0
		                               \end{pmatrix}
\end{equation}
where the momentum dependent mass matrix $\tilde M$ is diagonal in color space, 
\begin{equation*}
\tilde M=\text{diag}_\text{c}(M(\omega_n^-,\vec{p}\,),M(\omega_n^+,\vec{p}\,),M(\omega_n^0,\vec{p}\,))\,,
\end{equation*}
 with $\omega_n^\pm=\omega_n\pm A_4,\omega_n^0=\omega_n$.  The trace may be further simplified leading to
\begin{equation*}\label{omegasimpel}
	\Omega=-4 T\sum_{i=0,\pm} \sum_{n\in\Z}\int\dfrac{\diff^3 p}{(2\pi)^3}\,\ln\left[{\omega_n^i}^2+\vec{p}\,^2
+M^2(\omega_n^i,\vec{p}\,)\right]+\dfrac{\bar\sigma^2}{2G}+\mathcal{U}(\Phi,\Phi^*,T)\,.
	\tag{\ref{OmegaMatsubara}$'$}
\end{equation*}
This is the thermodynamic potential of the nonlocal PNJL model in mean field approximation\footnote{Diquark degrees of freedom are omitted in the present work and will be dealt with elsewhere.}.

\end{subsection}

\begin{subsection}{Gap equations and results}\label{gapequations}

Apart from the sigma field,  the nonlocal PNJL model includes two more degrees of freedom, $\Phi$ and $\Phi^*$ or $A_4^3$ and $A_4^8$, that mimic the thermodynamics of confinement and deconfinement. Therefore,  three gap equations are derived and solved for $\bar\sigma$, $\Phi$ and $\Phi^*$. The necessary conditions are given by the requirement of a stationary potential:
\begin{equation}\label{gaps}
	\dfrac{\partial\Omega}{\partial \bar\sigma}=\dfrac{\partial \Omega}{\partial A_4^3}=\dfrac{\partial \Omega}{\partial A_4^8}=0\,.
\end{equation}
Following Refs.~\cite{Simon1,Simon2}, we have $\Phi=\Phi^*$ in the mean field approximation and, consequently, $A_4^8=0$.

In Fig.~\ref{condPhicurrmass} we show the results for the temperature dependence of the chiral condensate and of the Polyakov loop using the parameters given in Tables \ref{parameterfix}, \ref{polyakovparameters}. This figure illustrates once more, as already demonstrated in Refs.~\cite{Simon1,Simon2}, the entanglement of chiral dynamics and Polyakov loop degrees of freedom, a characteristic feature of the PNJL approach. In the absence of a coupling between quark quasiparticles and Polyakov loop, and in the chiral limit, the second order chiral phase transition (for $N_\text{f}=2$ flavors) and the first order deconfinement transition (of pure gauge QCD) appear at very different critical temperatures ($T_\text{chiral}\approx110\,\text{MeV}$ for  the chiral phase transition and $T_0\approx 270\,\text{MeV}$ for deconfinement). The presence of quarks breaks the $Z(3)$ symmetry explicitly and turns the first order deconfinement phase transition into a continuous crossover. The quark coupling to the Polyakov loop moves this transition to lower temperature. At the same time the chiral transition (with explicit symmetry breaking by non-zero quark mass) turns into a crossover at an upward-shifted temperature, just so that both transitions nearly coincide at a common temperature $T_c\approx 200\,\text{MeV}$.

This symmetry breaking pattern seems also to be realized in recent lattice QCD results \cite{Cheng} where a common chiral and deconfinement transition temperature $T_c=(196\pm3)\,\text{MeV}$ is observed. Fig.~\ref{condPhicurrmass} shows these lattice data for orientation\footnote{A direct comparison with present nonlocal PNJL calculation is not possible since the lattice computation has been performed with $N_\text{f}=2+1$ flavors and a pion mass $m_\pi\simeq200\,\text{MeV}$.}.

\begin{figure}
\begin{center}
\includegraphics[width=.7\textwidth]{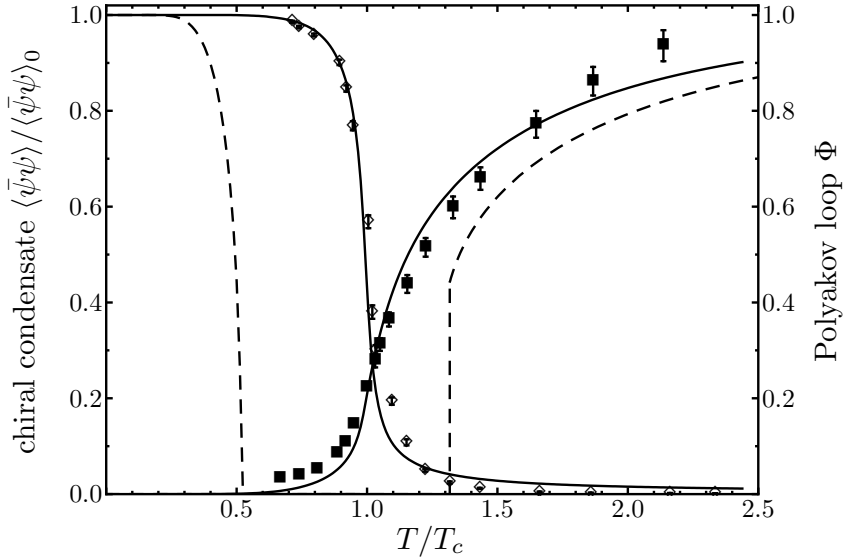}
\caption{Solid curves: calculated temperature dependence of the chiral condensate $\langle\bar\psi\psi\rangle$ and of the Polyakov loop $\Phi$ normalized to the critical temperature $T_c=207\,\text{MeV}$ as obtained in the nonlocal PNJL model. Also shown for orientation are lattice results from Ref.~\cite{Cheng}. The dashed lines show the chiral condensate in the chiral limit (left for the pure fermionic case) and of the Polyakov loop for the pure  gluonic case, respectively.}\label{condPhicurrmass}
\end{center}
\end{figure}

\end{subsection}

\begin{subsection}{Pressure and mesonic corrections}\label{press}

So far the calculations have been performed  in the mean field approximation in  which the pressure $P=-\Omega$ is determined by the quarks moving as quasiparticles in the background provided by the expectation values of the sigma field, $\bar\sigma$, and the Polyakov loop $\Phi$. In order to get a realistic description of the hadronic phase (at temperatures $T\lesssim T_c$), it is important to include mesonic correlations. The hadronic phase is dominated by pions as Goldstone bosons and their interactions. The quark-antiquark continuum is suppressed by confinement. However, as pointed out  in Ref.~\cite{Hansen}, mesons described within the standard (local) PNJL model can still undergo unphysical decays into the quark-antiquark continuum even below $T_c$. In the nonlocal PNJL model, those unphysical decays do not appear by virtue of the momentum dependent dynamical quark mass. Following Refs.~\cite{Bowler} one can easily show that, within our nonlocal NJL model, no real poles appear in the fermion propagator \eqref{ANJLfermion}  and hence mesons are stable well below $T_c$. This means that the pressure below $T_c$ is basically generated by the pion pole with its almost temperature independent position. Therefore the calculated pressure below $T_c$ corresponds to that of a boson gas with constant mass.

To include the mesonic contributions to the pressure in our nonlocal PNJL model we can basically use the formalism described in Ref.~\cite{Hansen}. One has to calculate the quark loop contribution to the mesonic self-energies $\varPi_{\pi,\sigma}(\nu_m,\vec{p}\,)$ (where $\nu_m=2\pi m T, m\in\Z$ is the bosonic Matsubara frequency and $\vec{p}$ is the momentum of the incoming pion or sigma), depicted in Eq.~\eqref{G}, at finite temperature. Using the rules \eqref{matsubarasubs} we end up with
\begin{equation}\label{tempselfen}\begin{aligned}
 	\varPi_{\pi,\sigma}(\nu_m,\vec{p}\,)&=8 T\sum_{i=0,\pm}\sum_{n\in\Z}\int\dfrac{\diff^3 k}{(2\pi)^3}\,\mathcal{C}(\omega_n^i+\nu_m,\vec{k}+\vec{p}\,)\,\mathcal{C}(\omega_n^i,\vec{k}\,)\times\\
&\times\dfrac{\omega_n^i(\omega_n^i+\nu_m)+\vec{k}(\vec{k}+\vec{p}\,)\pm M(\omega_n^i+\nu_m,\vec{k}+\vec{p}\,)M(\omega_n^i,\vec{k}\,)}{\left[(\omega_n^i+\nu_m)^2+(\vec{k}+\vec{p}\,)^2+M^2(\omega_n^i+\nu_m,\vec{k}+\vec{p}\,)\right]\left[(\omega_n^i)^2+\vec{k}\,^2+M^2(\omega_n^i,\vec{k}\,)\right]}\,,
	\end{aligned}
\end{equation}
where $\omega_n^\pm=\omega_n\pm A_4,\omega^0=\omega_n$ and $M(\omega_n,\vec{p}\,)=m_q+\mathcal{C}(\omega_n,\vec{p}\,)\bar\sigma$. The additional contribution of mesonic quark-antiquark modes to the pressure is given by a ring sum of RPA chains, investigated in Ref.~\cite{Huefner} and leading to the expression
\begin{equation}\label{mesonpress}
 	P_\text{meson}(T)=-T\sum_{M=\pi,\sigma}\dfrac{d_M}{2} \sum_{m\in \Z}\int\dfrac{\diff^3 p}{(2\pi)^3}\ln\left[1- G\varPi_M(\nu_m,\vec{p}\,)\right],
\end{equation}
where $d_M$ is the mesonic degeneracy factor, i.\,e.~$d_\pi=3,d_\sigma=1$.
Due to the momentum dependence of the nonlocality distribution $\mathcal{C}(p)$ and the dynamical quark mass $M(p)$, integrations and summations in Eqs.~\eqref{tempselfen} and \eqref{mesonpress} can only be carried out numerically. 

Results for the pressure in the presence of pion and sigma mesonic modes are presented in Fig.~\ref{pressplot}. Apart from the full result (solid line) we additionally show the mean-field result (with the pressure determined by quark quasiparticles only) and the mean-field result plus pion contributions. It is evident that at low temperatures the mean-field contribution stemming from the quarks is suppressed and the pressure can be described by a free pion gas. Near the critical temperature the sigma mesonic mode gives a small additional contribution. Finally, above temperatures $T>1.5\,T_c$ the mesonic contributions become negligible and the quark-gluon mean-fields dominate the pressure.

\begin{figure}
\begin{center}
\includegraphics[width=.7\textwidth]{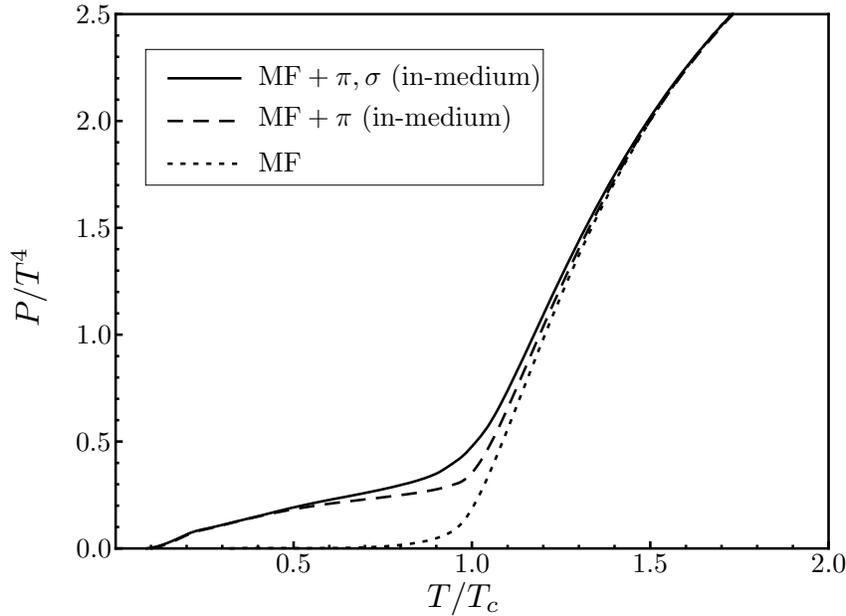}
\caption{Pressure calculated in the nonlocal PNJL model in units of $T^4$ as a function of  temperature normalized to the critical temperature. Solid curve: full calculation (i.\,e.~mean field result plus mesonic corrections). Dotted curve: mean-field result (no mesonic corrections). Dashed curve: mean-field plus pionic modes (no sigma).\label{pressplot}}
\end{center}
\end{figure}

Finally, in Fig.~\ref{comparisonpress} we show a comparison of the pressure calculated with the physical pion mass, $m_\pi=140\,\text{MeV}$, and with a ``heavy'' pion ($m_\pi=500\,\text{MeV}$) corresponding to a quark mass of order $m_q\sim100\,\text{MeV}$ that has frequently been used in earlier lattice QCD computations. In this case the mesonic contributions to the pressure are evidently reduced. This explains the apparent agreement of lattice data with mean field calculations \cite{Ratti1,Simon1}. 

\begin{figure}[t]
\begin{center}
		\includegraphics[width=0.7\textwidth]{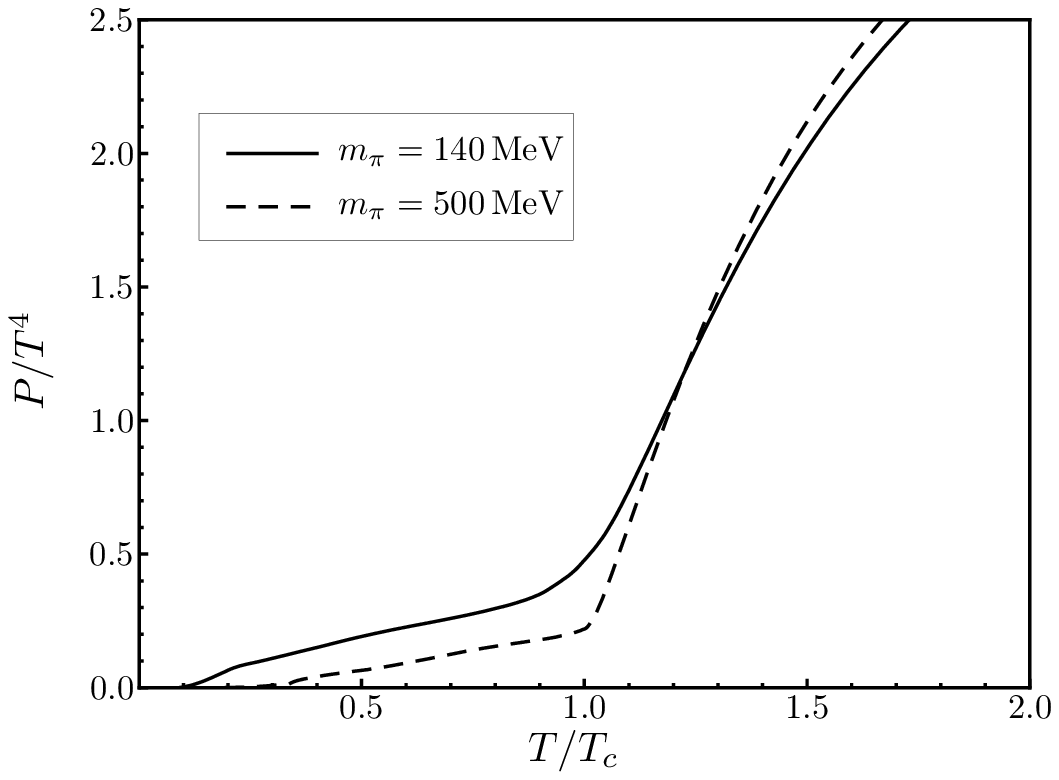}
		\caption{Comparison of the pressure for physical pion mass, $m_\pi=140\,\text{MeV}$ (solid line), and  heavy pion mass, $m_\pi=500\,\text{MeV}$ (dashed line). Note that the crossing of the curves is due to the different critical temperatures they were normalized to.\label{comparisonpress}}
\end{center}
\end{figure}

\end{subsection}

\begin{subsection}{Related thermodynamic quantities}
 
Given the partition function (or the thermodynamical potential $\Omega$) we can further investigate the energy density $\varepsilon$, the trace anomaly $(\varepsilon-3 P)/T^4$ and the sound velocity $v_\text{s}$ (compare also Refs.~\cite{Gosh}). In particular, the trace anomaly is of interest here since it is the  quantity which can be directly computed in lattice simulations (Ref.~\cite{Cheng}). The trace anomaly,  corresponding to the trace of the energy-momentum tensor, is the ``interaction measure'' which -- using thermodynamical relations -- can be expressed in terms of a derivative of the pressure with respect to temperature:
\begin{equation*}
 	\dfrac{\varepsilon-3 P}{T^4}=T\dfrac{\partial}{\partial T}\left(\dfrac{P}{T^4}\right).
\end{equation*}
A further interesting quantity is the ratio of pressure and energy density. Finally, the square of the sound velocity (at constant entropy $S$) is deduced as:
\begin{equation*}
 	v_\text{s}^2=\left.\dfrac{\partial P}{\partial \varepsilon}\right|_S=\dfrac{\left.\frac{\partial P}{\partial T}\right|_V}{T\left.\frac{\partial^2 P}{\partial T^2}\right|_V}\,,
\end{equation*}
which is connected to the second derivative of the thermodynamic potential.

The following figures  show the quantities just mentioned as they result in the mean field case and with inclusion of mesonic corrections. Again, mesonic corrections are  important  only at temperatures below $T_c$ (Figs.~\ref{conf}--\ref{vs}). Mesonic correlations do have a strong influence on the sound velocity below $T_c$ (Fig.~\ref{vs}). At this point it is not yet possible to compare these results to lattice data quantitatively since, as pointed out in the previous section, for larger meson masses, these corrections become less important and consequently their impact on $v_\text{s}$ diminishes.

\begin{figure}[t]
\begin{center}
	\begin{minipage}[t]{.475\textwidth}{
		\includegraphics[width=\textwidth]{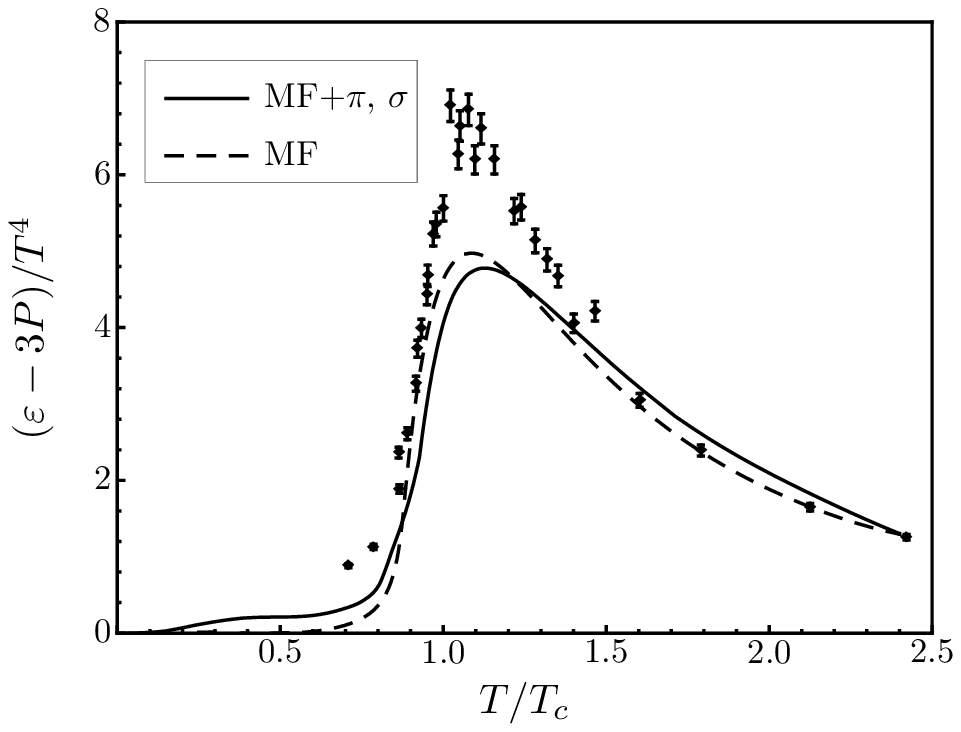}
		\caption{Trace anomaly $(\varepsilon-3 P)/T^4$ shown as a function of temperature for the mean field case and with mesonic correlations added. For orientation, lattice data are shown from Ref.~\cite{Cheng}. \label{conf}}}
	\end{minipage}
\hfill
	\begin{minipage}[t]{.475\textwidth}{
		\includegraphics[width=\textwidth]{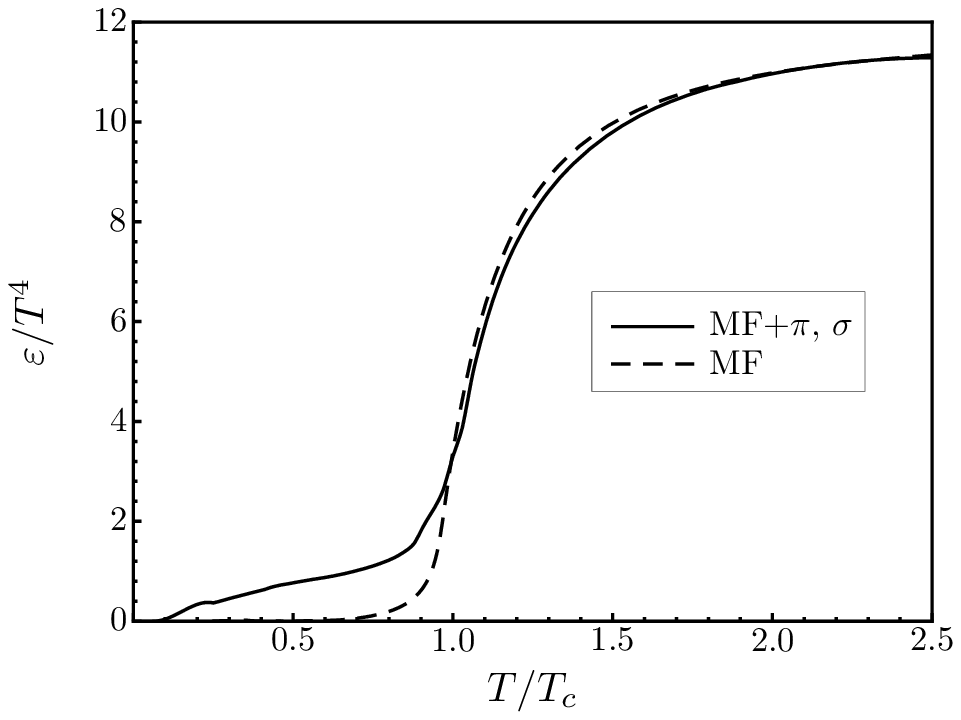}
		\caption{Comparison of the energy density $\varepsilon$ as a function of temperature for the mean field case and with mesonic correlations. \label{eps}}}
	\end{minipage}
\end{center}

\begin{center}
	\begin{minipage}[t]{.475\textwidth}{
		\includegraphics[width=\textwidth]{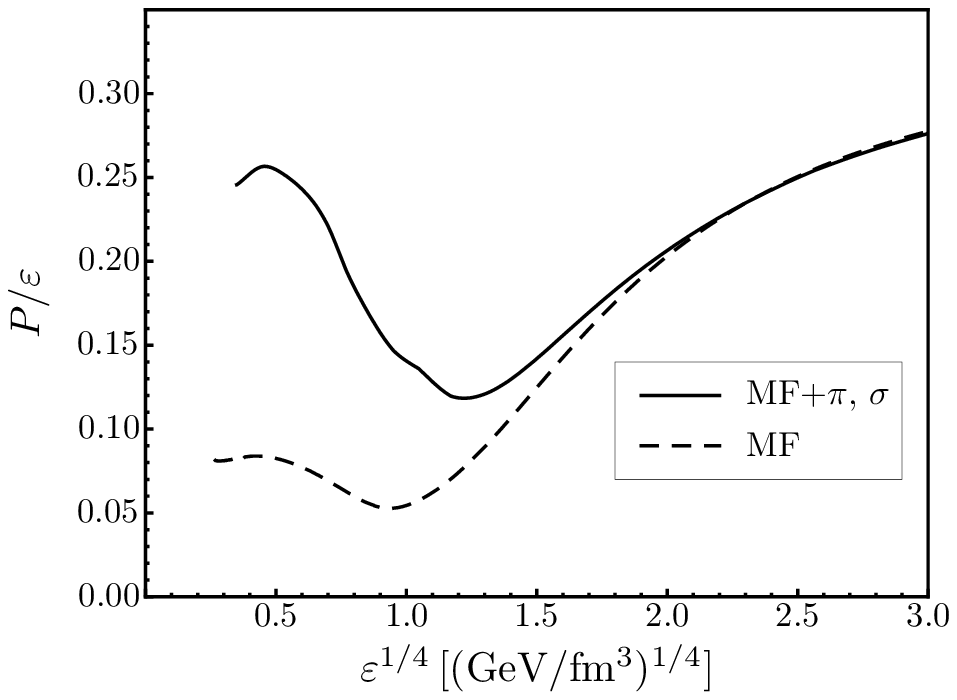}
		\caption{Fraction of  pressure and energy density $P/\varepsilon$ as a function of 4th root of the energy density $\varepsilon^{1/4}$.  \label{peps}}}
	\end{minipage}
\hfill
	\begin{minipage}[t]{.475\textwidth}{
		\includegraphics[width=\textwidth]{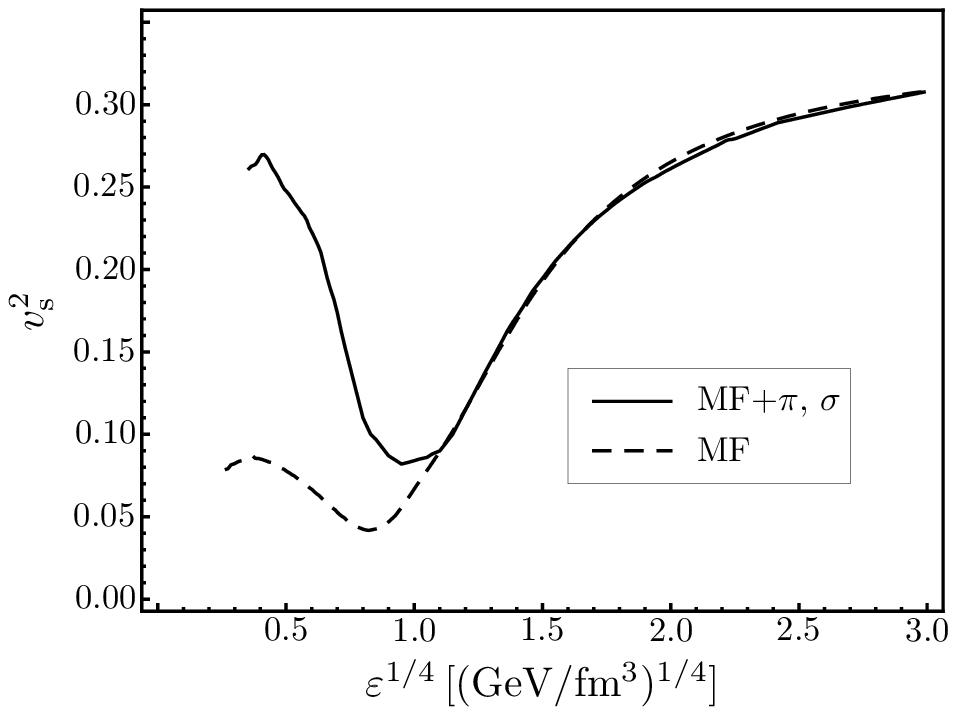}
		\caption{Squared sound velocity $v_\text{s}^2$ as a function of 4th root of the energy density $\varepsilon^{1/4}$. \label{vs}}}
	\end{minipage}
\end{center}
\end{figure}

\end{subsection}

\begin{subsection}{Finite quark chemical potential and phase diagram}

The nonlocal PNJL approach described in this work can be extended to finite quark chemical potential $\mu$. We do this here with the aim of drawing a schematic phase diagram in the $(T,\mu)$ plane. For the sake of simplicity we restrict ourselves to a scenario without diquarks. Their detailed treatment will be reported in a forthcoming paper. The introduction of a chemical potential\footnote{For convenience we use a quark chemical potential throughout this work. The corresponding baryon chemical potential is three times the quark chemical potential, i.\,e.~$\mu_\text{B}=3\mu$.} is accomplished using the  prescriptions of the Matsubara formalism (see Refs.~\cite{Kapusta}):  shift the frequencies $\omega_n\to\omega_n-\imu\mu$ in the particle sector (i.\,e. upper-left submatrix) of the Nambu-Gor'kov propagator \eqref{nambugorkovS} and  replace $\omega_n\to\omega_n+\imu\mu$ in the corresponding anti-particle sector (i.\,e. lower-right submatrix). It is then straightforward to compute the thermodynamic potential $\Omega(T,\mu)$ at non-zero $\mu$, following Eqs.~\eqref{OmegaMatsubara}, \eqref{nambugorkovS} with Matsubara frequencies properly shifted by the chemical potential.

We focus here on the $T$- and $\mu$-dependence of the scalar field $\bar\sigma$ acting as a chiral order parameter, deduced from the condition $\frac{\partial\Omega(T,\mu)}{\partial\bar\sigma}=0$. The result is shown in Fig.~\ref{phaseplot3d}. The profile of $\bar\sigma$ displays once again the chiral crossover transition at $\mu=0$. It turns into a first order phase transition at a critical point (here: $T_\text{CEP}=167\,\text{MeV}$ and $\mu_\text{CEP}=175\,\text{MeV}$). This qualitative feature is typical for NJL or PNJL type models with or without explicit diquark degrees of freedom (see e.\,g.~Refs.~\cite{Ratti1,Simon1,Buballa}). Related work is reported in Refs.~\cite{Sasaki,Abuki} where the nonlocality of the Fermionic interaction is introduced only in the three-momentum sector. The critical point found in these calculations differs both in its $T$- and $\mu$-value from the aforementioned references. Reasons for such discrepancies are given in Refs.~\cite{Simon1,Fukushima3} where it is pointed out and demonstrated that the location of the critical point is extremely sensitive to model details and input parameters.

Of course, the picture drawn in Fig.~\ref{phaseplot3d} is to be taken as only schematic, for several reasons. First, the location of the critical point is extremely sensitive to the input current quark mass \cite{Simon1}. Moreover, extensions to $N_\text{f}=2+1$ flavors including the axial $\text{U}(1)_\text{A}$ anomaly (Refs.~\cite{Fukushima3,Yamamoto}) point out uncercainties not only in the location of the critical point but also for the mere existence of a first-order phase transition. Secondly, the almost constant behavior of $\bar\sigma(T=0,\mu)$ with increasing chemical potential is unrealistic in the absence of explicit baryon (nucleon) degrees of freedom including their interactions.

What is actually required as a starting point for extensions to non-zero chemical potential is a realistic equation of state at finite baryon density, incorporating the known properties of equilibrium and compressed nuclear matter. In such a framework, the density dependence of the chiral condensate $\langle\bar\psi\psi\rangle$ (or of the scalar field $\bar\sigma$) is well known to be quite different from the profile shown in Fig.~\ref{phaseplot3d}. The magnitude of $\langle\bar\psi\psi\rangle$ decreases linearly with density $\rho$ \cite{densdep}, with a slope controlled by the pion-nucleon sigma term, and then stabilizes at densities above normal nuclear matter through a combination of two- and three-body correlations and Pauli blocking effects (see e.\,g.~Ref.~\cite{Kaiser}).

Irrespective of these comments, the nonlocal PNJL approach is obviously instructive in modeling the chiral and deconfinement thermodynamics at $\mu=0$. Dealing with finite baryon density requires ultimately yet another synthesis, namely that of PNJL and in-medium chiral effective field theory with baryons.

\begin{figure}[t]
\begin{center}
		\includegraphics[width=0.7\textwidth]{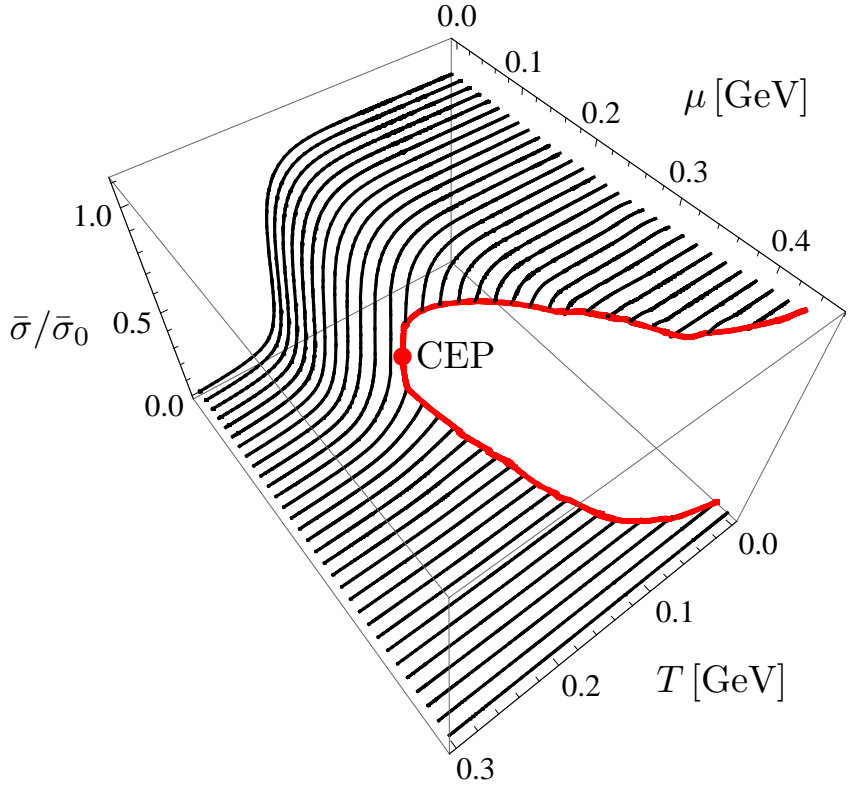}
		\caption{(Normalized) chiral order parameter $\bar\sigma/\bar\sigma_0$ shown in the $(T,\mu)$ plane. For small chemical potentials a crossover transition is manifest. For large chemical potentials a first order phase transition is apparent which terminates in the critical point (CEP) at $(T_\text{CEP},\mu_\text{CEP})=(167\,\text{MeV},175\,\text{MeV})$.\label{phaseplot3d}}
\end{center}
\end{figure}

\end{subsection}

\end{section}

\begin{section}{Summary and conclusions}\label{Conclusion}

A nonlocal generalization of the two-flavor PNJL model, a synthesis of Polyakov loop dynamics with the $N_\text{f}=2$ Nambu and Jona-Lasinio model, has been derived with the aim of identifying dominant quasiparticle deegrees of freedom and developing insight into the symmetry breaking scenario of QCD thermodynamics. This nonlocal approach has the principal advantage that it does not require the momentum space cutoff typical of NJL type models with local four-fermion couplings. The nonlocality of the interaction generates a momentum dependent dynamical quark mass, $M(p)$. Important QCD constraints can now be directly implemented,  e.\,g. through results from Landau gauge Dyson-Schwinger or lattice QCD calculations, or in contact with the instanton model, and in combination with the known high-momentum limit of $M(p)$. Remaining parameters are determined by reproducing pion properties in vacuum. Low-energy theorems based on chiral symmetry are shown to be rigorously fulfilled.
The $Z(3)$ symmetric effective potential governing Polyakov loop dynamics is fixed by comparison with the pure-gauge QCD equation of state from lattice QCD. Overall color neutrality is guaranteed by construction. 

The calculated chiral condensate and Polyakov loop as function of temperature show the dynamical entanglement of chiral and deconfinement transition which is characteristic of the PNJL approach: in the absence of quark-Polyakov loop coupling, the chiral phase transition (second order in the chiral limit) and the deconfinement phase transition (first order in pure gauge QCD) are separated in their critical temperatures by more than $150\,\text{MeV}$. Once the Polyakov loop couples to the quarks and suppresses their quasiparticle Green functions in  the confinement region, the two transitions (with explicit $Z(3)$ symmetry breaking by the presence of quarks and explicit chiral symmetry breaking by non-zero quark masses) become crossovers at a common transition temperature $T_c\simeq0.2\,\text{GeV}$. The present nonlocal PNJL approach demonstrates this as a general feature, without the need of introducing an artificial momentum cutoff.

Further results, such as the ``interaction measure'' $\varepsilon-3P$ and the sound velocity $v_\text{s}=\sqrt{\frac{\partial P}{\partial\varepsilon}}$, have been computed both at mean field level and with inclusion of mesonic (quark-antiquark) modes in the pion and ``sigma'' channels. As expected, a substantial contribution to the pressure and energy density below $T_c$ comes from propagating thermal pions. Around and above $T_c$ the mesonic modes quickly disappear and the thermodynamics is dominated by quark quasiparticles, with momentum dependent dynamical masses, and Polyakov loop degrees of freedom. As a schematic extension to non-zero quark chemical potential $\mu$, the chiral phase diagram in the $(T,\mu)$ plane has been  calculated and discussed.

The nonlocal PNJL model thus proves to be a good starting point for modeling QCD thermodynamics. It has the distinct advantage over previous (local) versions that the input (at the level of the dynamical quark mass function) can now be directly constrained by QCD.

Forthcoming work will be directed towards extensions with $N_\text{f}=2+1$ quark flavors such that direct comparisons can be performed with corresponding realistic lattice QCD computations.

\end{section}

\newpage
\begin{appendix}
\begin{section}{Bosonization of the nonlocal Lagrangian}\label{bosonizationApp}

 This appendix  briefly outlines how the bosonized form of the Euclidean action $\mathcal{S}_\text{E}^\text{bos}$, Eq.~\eqref{Sbos}, is derived from the action $\mathcal{S}_\text{E}$, Eq.~\eqref{S}. For this purpose we multiply 
\begin{equation}\label{genfunc}
		Z=\int\mathscr{D}\bar\psi\mathscr{D}\psi\,\euler^{-\mathcal{S}_\text{E}}
\end{equation}
by the constant $\mathcal{N}^{-1}:=\int\mathscr{D}\sigma\mathscr{D}\vec{\pi}\,\exp\left(-\int\diff^4 z\,\mathcal{C}(z)\int\,\diff^4 x\,\frac{\varphi_\alpha(x) \varphi_\alpha^*(x)}{2G }\right)$, with the mesonic fields $\varphi_a(x)=(\sigma(x),\vec{\pi}(x))$. The ``shift" $\varphi_\alpha(x)\to \varphi_\alpha(x)+G\, \bar\psi\!\left(x+\frac{z}{2}\right)\varGamma_\alpha\psi\!\left(x-\frac{z}{2}\right)$ allows then to ``complete the square" in the exponential of the generating functional and to eliminate the quadratic term in the nonlocal currents. Note that we have to  demand $\gamma_0\varGamma_\alpha^\dagger\gamma_0=\varGamma_\alpha$, as satisfied by the operators $\varGamma$ considered in this work. We end up with
\begin{equation*}
	\begin{aligned}
	\mathcal{S}_\text{E}&\to \int\diff^4 x\,\diff^4 y\,\bar\psi(x)\left[\delta^{(4)}(x-y)\left(-\imu\slashed{\partial}+m_q\right)+\mathcal{C}(x-y)\varGamma_\alpha\,\text{Re}\,\varphi_\alpha\!\left(\frac{x+y}{2}\right)\right]\psi(y)\\
	&\qquad +\dfrac{1}{2 G }\int\diff^4 z\,\mathcal{C}(z)\int\diff^4 x\, \varphi_\alpha(x)\,\varphi_\alpha^*(x)\,.
	\end{aligned}
\end{equation*}

The generating  functional \eqref{genfunc} can now be written as
\begin{equation*}\begin{aligned}
	Z&=\mathcal{N}\int\mathscr{D}\bar\psi\mathscr{D}\psi\mathscr{D}\sigma\mathscr{D}\vec{\pi}\\
		&\qquad\exp\left\{-\int\diff^4 x\,\diff^4 y\,\bar\psi(x)\left[\delta^{(4)}(x-y)\left(-\imu\slashed{\partial}+m_q\right)+\mathcal{C}(x-y)\varGamma_\alpha\,\text{Re}\,\varphi_\alpha\!\left(\frac{x+y}{2}\right)\right]\psi(y)\right\}\\
		&\!\!\!\!\qquad\times\exp\left\{-\dfrac{1}{2 G}\int\diff^4 z\,\mathcal{C}(z)\int\diff^4 x\, \varphi_\alpha(x)\,\varphi_\alpha^*(x)\right\}
\end{aligned}	\tag{\ref{genfunc}$'$}
\end{equation*}
and the integration over the fermionic degrees of freedom, $\bar\psi$ and $\psi$, can be performed, leading to
\begin{equation*}
	\begin{aligned}
Z&=\mathcal{N}\,\widetilde{\det}\left[\delta^{(4)}(x-y)\left(-\imu\slashed{\partial}+m_a\right)+\text{Re}\,\varphi_\alpha\!\left(\frac{x+y}{2}\right)\mathcal{C}(x-y)\varGamma_\alpha\right]\\
&\qquad\qquad\times\exp\left\{-\dfrac{1}{2 G }\int\diff^4 z\,\mathcal{C}(z)\int\diff^4 x\, \varphi_\alpha(x)\,\varphi_\alpha^*(x)\right\},
\end{aligned}
	\tag{\ref{genfunc}$''$}
\end{equation*}
where $\widetilde{\det}$ denotes the \emph{functional} determinant.

For the calculations done in this work it is more convenient to write integrals in momentum space than in coordinate space. Since a Fourier transform is a Hilbert space isomorphism, the integrals may be expressed in momentum space without further changes. A straightforward calculation leads to our final result
\begin{equation}\begin{aligned}
	Z&=\mathcal{N}\int \mathscr{D}\sigma\mathscr{D}\vec{\pi}\,\widetilde{\det}\left[(2\pi)^4\delta^{(4)}(p-p')\left(-\slashed{p}+m_q\right)+\frac{1}{2}(\phi_\alpha(p-p')+\phi_\alpha^*(p-p'))\varGamma_\alpha\,\mathcal{C}\!\left(\frac{p+p'}{2}\right)\right]\\
&\qquad\qquad\qquad\qquad\times\exp\left\{-\dfrac{1}{2G}\int\dfrac{\diff^4 p}{(2\pi)^4} \phi_\alpha(p)\phi_\alpha^*(p)\right\}.
	\end{aligned}
\end{equation}
From this generating functional the bosonized action can be read off:
\begin{equation}\begin{aligned}
	\mathcal{S}_\text{E}^\text{bos}=&-\ln\,\det\left[(2\pi)^4\delta^{(4)}(p-p')\left(-\slashed{p}+m_q\right)+\frac{1}{2}(\phi_\alpha(p-p')+\phi_\alpha^*(p-p'))\varGamma_\alpha \,\mathcal{C}\!\left(\frac{p+p'}{2}\right)\right]+\\
&+\dfrac{1}{2G}\int\dfrac{\diff^4 p}{(2\pi)^4} \phi_\alpha(p) \phi_\alpha^*(p)\,,
	\end{aligned}
\end{equation}
as indicated in Eq.~\eqref{Sbos}.

\end{section}

\newpage

\begin{section}{Taylor expansion of the Euclidean action \boldmath{$\mathcal{S}_\text{E}^\text{bos}$}}\label{taylor}

In this appendix we derive the Taylor expansion for the bosonized Euclidean action $\mathcal{S}_\text{E}^\text{bos}$ up to terms of second order. This corresponds to a systematic perturbative expansion and to the derivation of Feynman rules for the mesonic degrees of freedom according to the method of the ``effective action".

We start with a reminder of the Taylor expansion of a functional $T=T[f(x)]$ about  $f(0)$ with a (small) fluctuation $h(x)$. One has
\begin{equation}\begin{aligned}
	T[f(0)+h(x)]=T[f(0)]&+\int\diff y_1\left.\dfrac{\delta T[f(x)]}{\delta f(y_1)}\right|_{f=f(0)} h(y_1)+\\
&+\dfrac{1}{2}\int\diff y_1\,\diff y_2\left.\dfrac{\delta^2 T[f(x)]}{\delta f(y_2)\,\delta f(y_1)}\right|_{f=f(0)} h(y_1)h(y_1)+\dots
	\end{aligned}
\end{equation}

This formalism is now applied to $\mathcal{S}_\text{E}^\text{bos}$. In this case, $f(0)=(\bar\sigma,\vec{0}\,)$ and $h(x)=(\delta\sigma(x),\delta\vec{\pi}(x))$. The following relation is useful:
\begin{equation*}
	\ln\,\det=\text{Tr}\,\ln\,,
\end{equation*}
which holds both in functional and matrix space. We treat the functional space first, hence we decompose $\det=\widetilde{\det}\otimes\text{Det}$ or $\text{Tr}=\widetilde{\text{Tr}}\otimes\text{tr}$, where operators with a tilde act exclusively on functional and $\text{Det}, \text{tr}$ solely on Dirac-, flavor- and color space.

Then, the zeroth order term is easy to calculate since $\bar\sigma$ is a Lorentz invariant and, hence, proportional to unity in functional space, i.\,e.~$\bar\sigma\langle p'|p\rangle=\bar\sigma\frac{(2\pi)^4}{V^{(4)}}\delta^{(4)}(p-p')$. Therefore, the functional trace simply gives an integration over $p$ and the argument of the logarithm is given by $\text{Det}\left[-\slashed{p}+m_q+\mathcal{C}(p)\bar\sigma\right]=(p^2+M^2(p))^{2 N_\text{f} N_\text{c}}$. From this the mean field result  \eqref{SMF} is deduced. The linear term vanishes by definition of the mean fields.

In order to determine the term of second order we calculate $\left.\frac{\delta^2}{\delta\sigma(k)\,\delta\sigma(\ell)}\widetilde{\text{Tr}}\,\ln\hat A\right|_{\sigma(x)=\bar\sigma}$. As a continuation of matrix multiplication  the logarithm of an operator $\hat A$ is treated as a power series where the multiplication is given by the convolution. Therefore all matrix identities can be adopted in functional space, too, if all matrix products are replaced by convolutions. In particular, in this article we use  operators $\hat O=\hat O[f]$ that are supposed to fulfill  necessary convergence criteria so that the following identities hold:
\begin{equation}\begin{aligned}\label{operatorids}
	\dfrac{\delta}{\delta f(y)}\widetilde{\text{Tr}}\,\ln\left[\hat O[f(x)]\right]&=\widetilde{\text{Tr}}\left[\hat O^{-1}[f]\dfrac{\delta \hat O[f(x)]}{\delta f(y)}\right]\\
	\dfrac{\delta}{\delta f(y)}\widetilde{\text{Tr}}\left[\hat O^{-1}[f(x)]\right]&=-\widetilde{\text{Tr}}\left[\hat O^{-1}[f]\dfrac{\delta \hat O[f(x)]}{\delta f(y)}\hat O^{-1}[f]\right].
                 \end{aligned}
\end{equation}
In the case of the operator $\hat A$, Eq.~\eqref{A}, we may write
\begin{equation*}
\dfrac{\delta^2}{\delta\sigma(k)\,\delta\sigma(\ell)}\widetilde{\text{Tr}}\,\ln\hat A=-\widetilde{\text{Tr}}\left[(\mathcal{C}\,\delta^{(4)}(\bullet-\ell))\hat A^{-1}(\mathcal{C}\,\delta^{(4)}(\bullet-k)) \hat A^{-1}\right],
\end{equation*}
where the $\bullet$ stands for the arguments of the delta-function. Taking into account
\begin{equation*}
	\left.\langle p|\hat A|p'\rangle\right|_{(\sigma,\vec{\pi}\,)=(\bar\sigma,\vec{0})}=(2\pi)^4\delta^{(4)}(p-p')\left[-\slashed{p}+m_q+\mathcal{C}(p)\bar\sigma\right],
\end{equation*}
one  carries out the convolutions and arrives at the result:
\begin{equation*}\begin{aligned}
	\dfrac{\delta^2}{\delta\sigma(k)\,\delta\sigma(\ell)}\widetilde{\text{Tr}}\,\ln\hat A=-\dfrac{1}{(2\pi)^4}\,\delta^{(4)}(k+\ell)\,\mathcal{C}(k)\,\mathcal{C}(\ell)
\int\dfrac{\diff^4 p}{(2\pi)^4}&\left[-\slashed{p}-\slashed{\ell}+m_q+\mathcal{C}(p-\ell)\bar\sigma\right]^{-1}\\ &\times \left[-\slashed{p}+m_q+\mathcal{C}(p)\bar\sigma\right]^{-1}.
\end{aligned}
	\end{equation*}
Next, calculate the traces over Dirac-, flavor- and color space; moreover the expansion of the quadratic term in $\phi_\alpha$ is trivially done, leading altogether to the result stated in Eq.~\eqref{S2}.
\end{section}

\begin{section}{Pion decay constant}\label{decayconstant}

 Here we present the full expression for the pion decay constant defined as
	\begin{equation*}
	 	\langle0|J_{A,i}^\mu(0)|\tilde\pi_j(p)\rangle=\imu\,f_{\pi}\,\delta_{ij}\,p_\mu\,,
\tag{\ref{piondecaydef}}
	\end{equation*}
where $J_{A,i}^\mu$ denotes the axial-vector current.

In order to calculate the axial current matrix element one has to gauge the nonlocal action in Eq.\,\eqref{S}. This requires not only the replacement of the partial derivative by a covariant derivative,
	\begin{equation*}
	 	\partial_\mu\to\partial_\mu+\dfrac{\imu}{2}\gamma_5\tau_i\,\mathcal{A}_\mu^i(x),
	\end{equation*}
where $\tau_i$ are the Pauli matrices and $\mathcal{A}_\mu^i$  $(i\in\{1,\dots,3\})$ are axial gauge fields, but also the connection of nonlocal terms through a corresponding parallel transport with a Wilson line,
\begin{equation*}
 	\mathcal{W}(x,y)=\mathcal{P}\exp\left\{\dfrac{\imu}{2}\int_0^1\diff\alpha\,\gamma_5\tau_i\,\mathcal{A}_i^\mu(x+(y-x)\alpha)\,(y_\mu-x_\mu)\right\},
\end{equation*}
where we have chosen a straight line that connects the points $x$ and $y$. Exemplarily, this means that expressions of the form $\bar\psi(x-z)\,\mathcal{C}(z)\,\psi(z-y)$ in the action $\mathcal{S}_\text{E}$, Eq.\,\eqref{S}, have to be replaced by $\bar\psi(x-z)\,\mathcal{W}(x-z,z)\,\mathcal{C}(z)\,\mathcal{W}(z,z-y)\,\psi(z-y)$, which guarantees the (local) gauge invariance of the underlying Lagrangian. Following the steps described in Sect.\,\ref{ANJLconstructionSect}, one finds that the only term affected by the gauging is the fermion determinant $\hat{A}$, Eq.\,\eqref{A}, which then reads in coordinate space:
\begin{equation}\label{gaugefermiondet} 
		\begin{aligned}
 	A^\text{G}(x,y)&=\left(-\imu\,\slashed\partial_y+\dfrac{1}{2}\gamma_5\tau_i\,\slashed{\mathcal{A}}^i+ m_q\right)\,\delta(x-y)+\\
		&\quad+\mathcal{C}(x-y)\,\mathcal{W}\!\left(x,\frac{x+y}{2}\right)\,\varGamma_\alpha\,\varphi_\alpha\!\left(\frac{x+y}{2}\right)\,\mathcal{W}\!\left(\frac{x+y}{2},y\right),
	\end{aligned}
\end{equation}
where $\varGamma_\alpha$ stands either for $\varGamma_0:=1$ or $\varGamma_i=\imu\,\gamma_5\tau_i$, and $\varphi_\alpha$ accordingly for either the scalar field, $\sigma$, or a pseudoscalar field, $\pi_i$.

The desired matrix element then follows from the gauged fermion determinant according to 
	\begin{equation}
	 	\langle0|J_{A,i}^\mu(0)|\pi_j(p)\rangle=-\left.\dfrac{\delta^2\,\ln\,\det A^\text{G}}{\delta\pi_j(p)\,\delta\mathcal{A}_\mu^i(t)}\right|_{\mathcal{A}=0\atop t=0},
	\end{equation}
where we have used the unrenormalized pion field, $\pi_j=g_{\pi qq}\,\tilde{\pi}_j$.
The complete calculation is rather lengthy, but straightforward and uses the operator identities \eqref{operatorids} and
\begin{equation*}
 	\left.\dfrac{\delta\mathcal{W}(x,y)}{\delta\mathcal{A}_i^\nu(t)}\right|_{\mathcal{A}=0\atop t=0}=\dfrac{\imu}{2}\int_0^1\diff\alpha\,\gamma_5\tau_i\,\delta(x+(y-x)\alpha)\,(y_\nu-x_\nu)\ .
\end{equation*}

After a Fourier transform of the resulting expression one obtains
\begin{equation}\label{pionapp}
 	\begin{aligned}
 	 	\langle0|J_{A,i}^\mu(0)|\pi_j(p)\rangle&=2\imu\,\text{tr}\{\tau_i,\tau_j\}\,\widetilde{\text{Tr}}\left\{\int_0^1\diff\alpha\,q_\mu\dfrac{\diff\mathcal{C}(q)}{\diff q^2}\dfrac{M(q_\alpha^+)}{{q_\alpha^+}^2+M^2(q_\alpha^+)}\right\}+\\
	&+2\imu\,\text{tr}\{\tau_i,\tau_j\}\,\widetilde{\text{Tr}}\left\{\mathcal{C}(q)\dfrac{q_\mu^+ M(q^-)}{\big({q^+}^2+M^2(q^+)\big)\big({q^-}^2+M^2(q^-)\big)}\right\}+\\
	&+2\imu\,\bar\sigma\,\text{tr}\{\tau_i,\tau_j\}\times\\
		&\qquad\times\widetilde{\text{Tr}}\left\{\int_0^1\diff\alpha \,q_\mu\dfrac{\diff\mathcal{C}(q)}{\diff q^2}\,\mathcal{C}\!\left(\!q\!-\!\frac{p}{2}\alpha\!\right)\dfrac{q_\alpha^+\cdot q_\alpha^-\!+\!M(q_\alpha^+)M(q_\alpha^-)}{\big({q_\alpha^+}^2\!+\!M^2(q_\alpha^+)\big)\big({q_\alpha^-}^2\!+\!M^2(q_\alpha^-)\big)}\right\},
 	\end{aligned}
\end{equation}
with
\begin{equation}\label{palpha}
 	\begin{aligned}
 	 	q_\alpha^+&=q+\frac{p}{2}(1-\alpha),\qquad& q_\alpha^-&=q-\frac{p}{2}(1+\alpha)\\
		q^+&=q+\frac{p}{2},& q^-&=q-\frac{p}{2}\,.
 	\end{aligned}
\end{equation}

Now, the decay constant can be derived from the expression \eqref{pionapp} and the definition, Eq.\,\eqref{piondecaydef}, by contraction with $p^\mu$, hence
\begin{equation}
 	f_{\pi}=\imu\,p_\mu\langle0|J_{A,i}^\mu(0)|\pi_i(p)\rangle \dfrac{g_{\pi qq}^{-1}}{m_\pi^2}\,,
\end{equation}
evaluated at the pion pole $p^2=-m_\pi^2$.

\end{section}

\end{appendix}

\end{document}